\documentclass{aa}  
\usepackage{xcolor}
\usepackage{graphicx}
\usepackage{overpic}
\usepackage{txfonts}
\usepackage{lipsum}
\usepackage[leftcaption]{sidecap}
\usepackage{amsmath}
\usepackage{multicol}
\usepackage{subcaption}         
\usepackage{lscape}             
\usepackage{placeins}           
                                
\usepackage{hyperref}
\hypersetup{
    colorlinks=true,
    linkcolor=blue,
    filecolor=magenta,      
    urlcolor=blue,
    citecolor=blue
    }

\usepackage{amssymb}
\usepackage{amsfonts}
\usepackage{soul}
\usepackage[dvipsnames]{xcolor}
\usepackage{lineno}
\nolinenumbers

\def\t{\text}
\def\aCO{\alpha_\text{CO}}
\def\H2{H$_2$}
\def\MH2{\t{M}_\t{H2}}

\def\M*{M$_\star$}
\def\Msun{M$_\odot$}
\def\SFRten{\t{SFR}_\t{10}}
\def\SFR100{\t{SFR}_\t{100}}
\def\fmol{f_\text{mol}}
\def\tdep{\tau_\text{dep}}
\def\kms{km~s$^{-1}$}

\begin{document}
  
   \titlerunning{Resolved molecular gas and star formation in UGC~8179}
    \title{Resolved molecular gas and star formation in massive unquenched spirals}
   \subtitle{I. UGC~8179}

   \author{R. Cologni\inst{1,2},
            S. Flesch\inst{1},
            P. Salomé\inst{1},
            D. Le Borgne\inst{2},
            M. Boquien\inst{3},
	        J. Freundlich\inst{4},
            P. Guillard\inst{2}, 
            U. Lisenfeld\inst{5,6},
	        F. Combes\inst{1,7}
            \and L. Bouscasse\inst{8} 
        }
    \authorrunning{Cologni et al.}

   \institute{LUX, Observatoire de Paris, PSL Research University, Sorbonne Université, CNRS, 75014 Paris, France
            \and Sorbonne Université, CNRS, UMR 7095, Institut d’Astrophysique de Paris, 98bis bd Arago, 75014 Paris, France,\\
             \email{romane.cologni@obspm.fr}
              \and Université Côte d’Azur, Observatoire de la Côte d’Azur, CNRS, Laboratoire Lagrange, 06000, Nice, France
            \and Observatoire astronomique, Université de Strasbourg, CNRS UMR 7550, Strasbourg, France
            \and Departamento de F\'isica Téorica y del Cosmos, Universidad de Granada, 18071 Granada, Spain
            \and Instituto Carlos I de Física Téorica y Computacional, Facultad de Ciencias, 18071 Granada, Spain
            \and Collège de France, 11 place Marcelin Berthelot, 75231 Paris, France
            \and IRAM, 300 rue de la Piscine, 38400 Saint-Martin-d'Hères, France}

   \date{Received 9 December 2025 / Accepted 25 February 2026}

  \abstract
   {
   Recent studies have uncovered a rare population of supermassive (\M* $\gtrsim 10^{11}$ \Msun), yet actively star-forming spiral galaxies -- super spiral galaxies (SSGs) -- whose existence challenges classical mass- and environment-quenching scenarios. 
   We investigate the resolved star-forming and molecular-gas properties of the nearby SSG UGC~8179 ($z=0.052$, $\log(\t{\M*/\Msun})=11.62$) in order to assess whether its local star formation (SF) follows the same physical processes as those observed in typical star-forming main-sequence (SFMS) spirals.
   We combined the first spatially resolved CO(1–0) interferometric observations of an SSG with pixel-by-pixel SED fitting, based on archival UV–to–mid-IR imaging. Our $3\arcsec\times3\arcsec$ ($\sim25$ kpc$^2$) pixel maps provide spatially resolved measurements of stellar mass, star formation rate (SFR), and molecular gas surface densities across its extended disk.
   UGC~8179 hosts a massive rotating molecular gas reservoir of M$_\t{H2} = 1.02 \times 10^{10}$ \Msun, yielding a standard molecular gas fraction and a typical depletion time of $\sim 1$ Gyr in the region observed by NOEMA,  despite its extreme mass. We derive lower limits of $\log_{10}\fmol \geq -1.61  \pm 0.06$ and $\log_{10}\tdep \geq 8.82 \pm 0.13$ at the scale of the galaxy. The large spatial extent of UGC~8179 enables us to probe low surface-density regimes hardly accessible in nearby disks ($\Sigma_\star < 10^7$ \Msun~kpc$^{-2}$; $\Sigma_{\rm SFR}<10^{-3.5}$ \Msun~yr$^{-1}$~kpc$^{-2}$). All three resolved scaling relations (resolved SFMS -- rSFMS, resolved Kennicutt Schmidt -- rKS, and resolved molecular gas main sequence -- rMGMS) are well defined. The rKS slope ($0.87\pm0.09$) is broadly consistent with unity, indicating standard local SF processes. The rSFMS shows a shallower global slope ($0.80\pm0.02$) due to a central suppression in specific SFR ($\sim - 0.5 $ dex), but a two-component fit restores agreement with literature relations at $\Sigma_\star\lesssim10^{7.2}$ \Msun~kpc$^{-2}$. This break suggests the influence of a bulge -- and possibly a bar -- driving a transition to a more dynamically regulated SF regime in the inner disk. A similar flattening in the rMGMS supports this interpretation. 
   UGC~8179 provides evidence that SSGs can sustain standard local SF processes while exhibiting central dynamical regulation at high stellar surface densities. Extending this analysis to our full sample of 19 SSGs and nearby massive unquenched spirals will enable us to test whether such regulation is a common feature among massive unquenched spirals. 
   }

   \keywords{galaxies: evolution --
            galaxies: ISM --
            galaxies: spirals --
            galaxies: star formation --
            galaxies: individual: UGC~8179
               }

   \maketitle

\nolinenumbers

\section{Introduction}

The overwhelming majority of massive galaxies (\M* $> 10^{11}$ \Msun) in the local universe are red ellipticals with little to no star formation (SF), i.e., gas-poor, dispersion-dominated spheroids \citep{Conroy&Wechsler09,Peng10,Cappellari16}. The apparent bias of high-mass galaxies toward passive early-type galaxies is strongly connected to the evolutionary stages that lower-mass galaxies experience when building their stellar mass. Galaxies are seldom isolated systems: in fact, they are likely to experience various types of interactions throughout their lifetime, which can lead to environmental quenching. Dynamical disruptions can arise either from high-speed flyby encounters -- causing tidal stripping -- or from ram-pressure stripping if the galaxy cruises in the intracluster medium, which rips off its gaseous material \citep[e.g.,][]{GunnGott72,Moore96, George25}. Simulations have shown that mergers -- either minor ones from satellites infalling onto the main galaxy or major ones from similar systems within a group or a cluster -- can cause disruptions that echo in the global morphological evolution of the galaxy \citep[e.g., bar emergence, bulge growth, disk destruction;][]{Hopkins09}. These morphological features are related to a more efficient gas-fueling of the central supermassive black hole (SMBH), subsequently more likely to trigger an active galactic nucleus (AGN) phase \citep{Ellison11,Choi24} yielding mostly negative feedback, shutting down the SF (i.e., quenching). In addition to AGN-feedback, another widely acknowledged mass-quenching process is virial shock heating, namely the virial shock undergone by the intergalactic gas infalling onto galaxies embedded in massive halos \citep[log$_{10}$~M$_\t{halo} \gtrsim 12$,][]{Birnboim&dekel03,Gabor&Dave15}. It creates a large hot gas reservoir in the halo that fails to cool and fuel the SF \citep[e.g.,][]{Cattaneo06}, especially if the AGN maintains gas heating. These processes (i) shut down SF in the galaxy -- either by destabilizing the gas through heating or shocks, making it unsuitable for SF, or by depriving the host of its gas content, tearing it off or ejecting it at kiloparsec-scale distances through powerful outflows \citep{Cicone14, Esposito24} -- and (ii) greatly disturb, or even destroy the rotationally supported disk of spiral galaxies \citep{Bournaud11}. 

However, a recent study by \citet{Ogle16} shows that at $z<0.3$, 6\% of the most luminous galaxies in the SDSS $r$ band are extremely massive spirals, the so-called super spiral galaxies (SSGs). They produced the Ogle Galaxy Catalog (OGC), which includes 84 SSGs \citep{Ogle19b}. Exhibiting stellar masses (\M*) exceeding 10$^{11}$ \Msun, large optical diameters of 55--140~kpc, and remarkable star formation rates (SFRs) ranging 2-32~\Msun/yr, they appear as gigantic unquenched spiral galaxies. Their evolutionary pathway differs from the bulk of galaxies in this mass range, they thus challenge by essence classical mass-quenching scenarios \citep{Peng10}. Supermassive spiral galaxies might be the most efficient at accumulating baryons and converting them into stars \citep{Posti19,Ogle19a}. Indeed, they exhibit high baryonic fractions $f_M$ = \M*/M$_\t{halo}$, approaching the cosmological baryon fraction of $f_b = \Omega_b/\Omega_c = 0.188$ \citep{planckcollaboration2020}. This result is much higher than what is expected from the reference stellar-to-halo mass ratio (SHMR) peaking at  $f_M \simeq 0.2 f_b$ \citep[for $L^*$,][]{Moster13}. When limiting their scope to spirals only, studies show a monotonous increase in the baryonic fraction ratio $f_\star = f_M/f_b$ with \M*, rather than the expected steep decline, reaching nearly 100\% efficiency, (i.e., no missing baryons) despite $\log$ M$_\t{halo} \gtrsim 12$ \citep{Posti19, DiTeodoro23}. This self-similarity in disks has also been found when exploring the HI-to-halo mass ratio \citep{Korsaga23}.

When looking for SSGs (\M* > 10$^{11.4}$ \Msun) in Horizon-AGN simulations, \citet{Jackson20} find that the progenitors of these outliers are either preserved disks with an unusual quiet merger history (30\%) or, for a majority of them (70\%), rejuvenated disks following a merger (mass ratio > 1:10) between an elliptical galaxy and a gas-rich satellite. This was a result confirmed by the observational follow-up study of \citet{Jackson22} using UV-optical (SDSS, GALEX, and DECaLS) and HI (ALFALFA) surveys. Additionally, \citet{Pallero25} reveal that supermassive disks are found in all environments in IllustrisTNG-100 simulations; however, they are prevalently in low-mass groups or isolation ($\sim$ 60\%) and generally have a quiet merger history. Thus, SSGs managed to build their stellar mass without destroying their disk and maintaining an active stellar formation, implying a replenished gas reservoir.

To better understand and characterize SF mechanisms in such galaxies, it is necessary to probe the interstellar medium (ISM) phase where SF takes place, namely their molecular gas. Unresolved detections of CO(1-0) in 46 SSGs with IRAM 30~m telescope point toward high molecular gas mass fractions ($\log\fmol = \log(\MH2$/\M*)$=-1.36 \pm 0.02$) and slightly higher molecular depletion times ($\log\tdep = \log(\MH2$/SFR) $=9.3\pm0.03$) than the lower-mass spirals located on the star-forming main sequence \citep[SFMS, $\log\tdep =9.00\pm0.02$][L23 henceforth]{Lisenfeld23}. To study the molecular gas morphology, dynamics, and properties more in depth, we built a sample of 19 targets distributed in two subsamples -- six SSGs from OGC and 13 baryon-rich, slightly less massive nearby spirals chosen from \citet{Posti19} and \citet{DiTeodoro23} -- and observed them in CO(1-0) with the Institut de Radioastronomie Millimétrique (IRAM) \textit{Northern Extended Millimetre Array} (NOEMA) interferometer to map the cold gas reservoir for the first time (Cologni et al. in prep.). NOEMA is the leading radio interferometer in the northern hemisphere, with recently upgraded receivers, and the most suited to survey CO emission in our sample of northern targets. These new interferometric data give us access to the cold gas dynamics, allowing us to explore rotational velocities and disk stability.

One key question is to investigate whether SSGs behave like SFMS standard spirals or if they obey different scaling laws. With this study, we aim at mapping the molecular gas, investigating the resolved star-forming properties ($\fmol$ and $\tdep$) of these extreme objects in a resolved manner, and obtaining more detailed insight into the physics of the SF taking place. For this purpose, we developed a method to model the spectral energy distributions (SEDs) of the subregions in the disk using the panchromatic SED fitting code CIGALE \citep{Boquien19}. Our goal is then to compare these results to the properties of standard spiral galaxies, located on the SFMS.

In this paper, we focus on the methodology and on applying it to one outstanding galaxy in the sample: UGC8179. The rest of the sample will be presented in an upcoming paper (Cologni et al., in prep.). To constrain the fits, we used available archival photometric data of GALEX in far- and near-UV bands, SDSS in optical $u$, $g$, $r$, $i$, and $z$ bands, together with WISE fluxes at 3.4, 4.6, and 12~µm wavelengths. Additionally, UGC 8179 has available integral field unit (IFU) observations from the MaNGA survey \citep{bundy2015} and was observed in HI with the Very Large Array (VLA) \citep{DiTeodoro23}.

This paper is organized as follows. Section~\ref{sec:observations} describes the observations as well as the reduction of the interferometric data from NOEMA (Sec.~\ref{subsec:molgas}) and photometric data from SDSS, GALEX, and WISE (Sec.~\ref{subsec:photo_data}). Section~\ref{sec:methods} explains the methodology used to estimate the global and local physical properties of UGC~8179 via SED modeling. Section~\ref{sec:results} presents the main results of the paper, namely the molecular gas detection (Sec.~\ref{subsec:CO}) and the global and resolved properties of UGC~8179 (Sec.~\ref{subsec:global_prop} and Sec.~\ref{subsec:resolved_prop}). These results are discussed in Section~\ref{sec:Discussion}, and the conclusions for this study are given in Section~\ref{sec:conclusions}. All rest-frame and derived quantities in this work assume a \citet{Chabrier03} initial mass function (IMF) and a cosmology with H$_0$ = 70.4 km s$^{-1}$ Mpc$^{-1}$, $\Omega_\text{m}$ = 0.272, and $\Omega_\Lambda =$ 0.728 \citep[Table 1 in ][WMAP + BAO + H$_0$ Maximum Likelihood parameters]{Komatsu11}. This results in a scale of 1.007~kpc/\arcsec and a luminosity distance of $D_L = 230$~Mpc.

\begin{table*}[!htbp]
   \begin{center}
   \centering
   \caption[]{Information on source UGC~8179.}
   \begin{tabular}{lcccccccccc}
    \hline\hline \noalign {\smallskip}
     source & z & RA & DEC & $i$ & Diam. & $R_e$\tablefootmark{a} & $D_L$ & SFR\tablefootmark{b} & log$_{10}$ M$_\star$/M$_\odot$\tablefootmark{c} & $E(B-V)$\\
     & & & & {\small deg} & {\small kpc} & {\small kpc} & {\small Mpc} & {\small M$_\odot$/yr} & {\small log$_{10}$(M$_\odot$)} & {\small mag}\\
      \hline \noalign {\smallskip}
       UGC~8179 & 0.051884 &13:05:14.15 &+31:59:59.0 & 65 & 122.4 & 19.0 & 230 & 17.8 & 11.4 $\pm$ 0.01& 0.0112 \\ 
    \hline \noalign {\smallskip}
    \end{tabular}
    \tablefoot{References:
    \tablefoottext{a}{Half-light radius r$_{50}$ from NASA Sloan Atlas, NSA ID: 090504.}
    \tablefoottext{b}{From UV and IR photometry, \citet{Lisenfeld23}.}
    \tablefoottext{c}{From WISE1 aperture photometry, \citet{DiTeodoro23}}
    }
    \label{tab:IDcard}
    \end{center}
\end{table*}

\section{Observations}
\label{sec:observations}
\subsection{UGC~8179}
The properties of UGC~8179 are utterly unusual for a spiral galaxy at $z = 0.0519$. With a stellar mass of $\log_{10}$ \M*$=11.4$ and SFR of $17.8$ \Msun~yr$^{-1}$, it is more similar to our OGC subsample ($z_\t{med}^\t{SSG}\sim0.17$), while being at a much closer distance that ensures better spatial resolution. Indeed, its vast disk spans a diameter > 120~kpc and $\sim$ 2\arcmin. UGC~8179 is part of a group of six galaxies \citep[group ID : \#02587,][]{Tempel17}, itself located in a populated environment. The number of galaxies within a 5 (10) Mpc and 2500 (5000) $\t{\kms}$ radial distance to UGC~8179 equals 67 (429). The group has not been shown to be merging with other groups, and the distances between the members within the group range 100 to 570~kpc. General information about UGC~8179 is summarized in Table~\ref{tab:IDcard}.

\subsection{Molecular gas data}
\label{subsec:molgas}

\subsubsection{CO observations with NOEMA interferometer}
The observation was carried out with the NOEMA interferometer. We retrieved data for UGC~8179 on three tracks between July 7 and 10, 2023, for a total on-source integration time of 3~h. The first two tracks used ten out of the 12 antennas in the compact D configuration (1.5~h on source), and during the last track nine antennas were active in D configuration (1.5~h on source). The single pointing aimed at the center of the galaxy, where we observed the $^{12}$CO(1-0) redshifted line at 109.584~GHz, considering $z = 0.051884$. The half-power beam width (HPBW $\approx 1.13 \lambda/D$) for CO(1-0) at this frequency equals 42.5\arcsec. The phase calibrator was J1310+323. The bandpass calibration was done on 3C345 and the source flux calibrated with MWC349, 3C345 and J1310+323.

NOEMA is equipped with a PolyFix correlator, processing a total instantaneous bandwidth of 15.5~GHz per polarization, distributed in two sidebands separated by $\sim15.5$~GHz. The spectral resolution is 2~MHz throughout the effective bandwidth. Data calibration was performed using the standard pipeline in \texttt{CLIC}, a software from the GILDAS\footnote{\url{http://www.iram.fr/IRAMFR/GILDAS}} package. Given the satisfactory signal-to-noise ratio (S/N) after data inspection, we ran the pipeline again to improve the data quality, choosing the channel interpolation method instead of fitting polynomials for RF calibration, chosen by default for two of the three tracks. The following steps of imaging and cleaning were performed using the software \texttt{MAPPING} of GILDAS package.

The rms is 0.566~mJy/beam at a spectral resolution of $\Delta w_\text{ch} = 21$~km~s$^{-1}$, and for a synthesized beam of $4.9\arcsec\times2.9\arcsec$, PA of 49 degrees from positive declination direction. We smoothed the spectrum of native resolution of 5~km~s$^{-1}$ (compression by a factor of 4). The resulting spectrum integrated over the entire detected CO disk is shown in the top panel of Figure \ref{fig:COmomentmaps}.

All data, including spectra, moment maps and channel maps, were made available online to allow one to reproduce the figures. The following link\footnote{\url{https://yafits-ssg.obspm.fr/summary.html}} give access to cleaned cube, corrected\footnote{s23bp\_id1\_comp4\_cleancube\_prim.json} or not\footnote{s23bp\_id1\_comp4\_cleancube.json} for the primary beam, as the data are intrinsically noisier at the edge of the field of view (FoV). The interfaces were created using Yafits \citep{Salome24}.

\subsubsection{Mapped molecular gas mass}
\label{subsubsec:conv_CO_H2}
In order to compute the molecular mass from the integrated CO flux density, we adopted equation (3) of \citet{Solomon97}:
\begin{equation}
    L^\prime_\text{CO}=3.25\times 10^7 S_\text{CO} \, \Delta V \; \nu^{-2}_\text{obs} \; D_L^2 \; (1+z)^{-3},
    \label{eq:COline}
\end{equation}
with the line luminosity $L^\prime_\text{CO}$ expressed in K~km~s$^{-1}$~pc$^{-2}$, $S_\text{CO} \, \Delta V $ the velocity-integrated line flux in Jy~km~s$^{-1}$, $D_L$ the luminosity distance in Mpc, $\nu_\text{obs}$ the redshifted frequency in GHz, and $z$ the redshift. Hence, we were able to deduce the molecular gas map in solar masses following:
\begin{equation}
    M_\text{mol} = \alpha_\text{CO} \times L^\prime_\text{CO},
    \label{eq:COtoH2}
\end{equation}
where $\aCO$ is the molecular gas mass-to-light ratio in M$_\odot$~(K~km~s$^{-1}$~pc$^{2}$)$^{-1}$, the conversion factor allowing one to infer H$_2$ mass from CO-line luminosity. This CO-to-H$_2$ conversion factor is known to depend on various properties, such as gas-phase metallicity or stellar surface density \citep{Accurso17,Genzel12,Schruba12,Bolatto13,Bisbas25}, resulting in a varying $\alpha_\text{CO}$ among and within galaxies. Thus, the assumption of a constant factor might seem too simplistic, especially for extreme sources such as SSGs. We investigated the possibility of a varying $\aCO$ using the local metallicity measurement from MaNGA data, and fitted a radial profile to it (methods are described in more detail in Appendix \ref{app:MaNGA_SFR}, see Fig.~\ref{fig:metallicity_radprofile_MaNGA}). We followed the formula based on calibrations provided by \citet{schinnerer2024} using \citet{hu2022} and \citet{chiang2024}: 
\begin{equation}
    \aCO^\t{MaNGA}(r) = 4.35 \: M_\odot \;(\t{K~km~s}^{-1}~\t{pc}^{2})^{-1} \times Z'(r)^{-1.5},
    \label{eq:alphaMaNGA}
\end{equation}
where $Z' = Z/Z_\odot = 10^{12+\log(O/H)-8.69}$ is the metallicity traced by [O/H] \citep[adopting $Z_\odot = 10^{12+\log\t{(O/H)}} = 10^{8.69}$][]{asplund2009}. We neglected the ``starburst'' term that depends on the stellar mass surface density. The H$_2$ masses presented in Section~\ref{sec:results} are obtained with this recipe, using the radial linear fit from the metallicity values computed with the more conservative methodology of \citet{schaefer2022}. Other values of $\alpha_\text{CO}$ and the subsequent influence on our results is explored in Section~\ref{subsec:discussion:alphaCO}.

The error bar on the H$_2$ mass is deduced from the error estimation of the CO(1-0) line. One can estimate the uncertainty the rms noise $\sigma_{\text{CO},\:\Delta W}$ in Jy in the detection region for the integrated linewidth $\Delta W$ (full width at zero intensity, FWZI), from the rms noise $\sigma_\text{CO, ch}$ in channels of width $\Delta w_\text{ch} = 21$~km~s$^{-1}$, around the CO(1-0) line: $\sigma_{\text{CO},\Delta W} = \sigma_\text{CO, ch}\:.\:\sqrt{\Delta w_\text{ch}/\Delta W\:}$ as commonly done \citep[e.g.,][]{Saintonge17}. We computed the uncertainty on the H$_2$ mass by integrating the rms noise $\sigma_{\text{CO},\:\Delta W}=0.62$~mJy, on the linewidth of $\Delta W = 875$~km~s$^{-1}$ and using eq. \ref{eq:COline}, we find $\sigma_\t{H2}=3.0\times10^{8}$ \Msun.

\subsection{Photometric data}
\label{subsec:photo_data}

\noindent\paragraph{SDSS.} The optical images were obtained thanks to the Sloan Digital Sky Survey Data Release 18 archive \citep[SDSS-V,][]{Kollmeier26}. The analysis was carried out on the available \textit{corrected frames}, which are the final step of processing, in the five $u$, $g$, $r$, $i$, and $z$ bands. The background subtraction was performed by subtracting the median value of the signal in an area greater than the regions considered.  
\vspace{-5mm}
\noindent\paragraph{GALEX.} Ultraviolet photometric measurements were made on tiles from the Galaxy Evolution Explorer \citep[GALEX,][]{Martin05}, retrieved on the Mikulski Archive for Space Telescopes GALEX GR6/7 archive. We downloaded both the near- and far-ultraviolet (NUV and FUV bands, respectively) intensity maps (\texttt{-int.fits}), in counts pix$^{-1}$~s$^{-1}$. The background subtraction was performed similarly to the SDSS images. 
\vspace{-5mm}
\noindent\paragraph{WISE.} Mid- to far-IR photometric measurements come from the WISE band 1, 2, 3, and 4 retrieved from the Wide-field Infrared Survey Explorer \citep[WISE,][]{Wright10} AllWISE Source Catalog, together with the uncertainty maps used in the error estimation (equation \ref{eq:err_WISE}).

\noindent Conversion to Jansky is detailed for all bands in Appendix \ref{app:Jyconversion}.

\vspace{1mm}
Spectral energy distribution modeling requires the extracted flux of the subregions, together with the estimated photometric errors, both in milliJansky. The computation of our estimate of the uncertainty for the considered fluxes is detailed in App. \ref{app:photo_err}. Keeping in mind that these estimations are a lower limit of the photometric flux uncertainty as they do not encompass all instrumental effects and systematic errors. We also added an error of 10\% of the flux in  quadratic sum for all the bands considered in this work, in order to take into account the neglected flux uncertainties as well as the modeling uncertainties, for example, due to model aliasing.

\subsection{Galactic extinction }

The reddening in the NIR being negligible, we only corrected the GALEX and SDSS data for Galactic extinction. We retrieved the Galactic dust reddening on the line of sight toward UGC~8179 from NASA/IPAC Infrared Science Archive\footnote{\url{https://irsa.ipac.caltech.edu/applications/DUST/}}. We used the $E(B-V) = 0.0112$~mag mean value, determined by \citet{Schlafly&Finkbeiner11}, who applied recalibration corrections to the prescriptions of \citet{SFD98}. This value was used to compute the extinction $A_\lambda = R_\lambda \times E(B-V)$, where $R_\lambda$ is defined for every band ranging from $FUV$ to $z$ band. The extinction curve in the UV (optical) range is computed following \citet{FM90, Fitzpatrick07}. The fluxes were thus corrected with $F_{\lambda,corr}=F_{\lambda} \times 10^{0.4A_\lambda}$. The corresponding values are listed in Table \ref{tab:Gal_extinction}. 

\section{SED fitting method}
\label{sec:methods}
 
A variety of photometric prescriptions for SFR exist in literature using UV light and mid-IR emission to account both for hot, newly born stars, and the warm dust continuum emission, excited by these young stars \citep[e.g.,][]{Hao11,Boquien16,Janowiecki17,Cluver17}. The total stellar mass \M* is generally derived from near-infrared light (e.g., at 3.6~$\mu$m from \textit{Spitzer}/IRAC or 3.4~$\mu$m from WISE1), which is dominated by the emission of old stars, and calculate \M* assuming a constant stellar mass-to-light ratio $\Upsilon_\star$ \citep[e.g.,][]{Eskew12}. 
However, this method has limitations: the assumption of a constant mass-to-light ratio overestimates \M* by $\sim$0.3-0.5 dex relative to SED-fitting results, especially since the deviation seems to depend on \M* \citep{Hunt19}, and our focus is on high-mass galaxies. \citet{Ogle16} and L23 rely on these recipes to estimate the global SFR and \M* of SSGs, and their sanity checks with global SED fitting point toward an overall good agreement. 

Most of these single- or dual-band SFR and \M* prescriptions were established using SED fitting \citep[e.g.,][]{Chang15} and rely on various assumptions and specific models that may not apply to SSGs. Hence, we chose to straightforwardly apply the SED fitting method to UGC~8179. In this way, we ensure a refined computation of SFR and \M* tailored on spatially resolved properties. There exist many different SED fitting codes based on various assumptions and code architectures, which subsequently result in a dispersion of estimates for a given property, as shown by \citet{Pacifici23}. We chose to use CIGALE \citep[Code Investigating GALaxy Emission][]{Boquien19}. It relies on energy balance between the absorption of UV to visible light by dust, and its reemission in the mid- and far-IR. One significant asset of CIGALE is its versatility: the diversity of modules used to build the library of models for the SED fitting allows easy tailoring of the parameter-space one wishes to explore. 

In this section, we discuss the motivation behind the modules used and the ranges of values adopted for the free parameters therein. Moreover, we describe the data preprocessing used to perform resolved SED fitting. For a more comprehensive description of the chosen modules, see \citet{Boquien19}. 

\subsection{SED fitting models}
\label{sec:modules}

CIGALE follows the standard steps when building models: computation of the SFH and of the unattenuated stellar spectrum from single stellar populations (SSPs), followed by computation of the lines and continuum nebular emission from the Lyman continuum photons emitted. Next, dust attenuation is applied to the nebular and stellar emissions, and the absorbed luminosity is used to compute the mid- and far-IR dust emission following the energy-balance principle. Lastly, redshifting the model and absorption by the intergalactic medium are taken into account. The code has a variety of modules for each physical component. 

We chose a classical delayed SFH with an optional constant burst or quench recent event (\texttt{sfhdelayed\_bq} module). We explored a wide range of parameters for both the age of the burst or quench episode (from 10 to 500~Myrs in look-back time) and the after/before ratio (spanning more than two orders of magnitude). We assumed the age of the first-born stars to be 12.7~Gyrs, corresponding to a birth around the epoch of reionization ($z\sim 6$).
Regarding the stellar continuum, we used SSPs of \citet{BruzualCharlot03} and restricted the values of stellar metallicity to 0.008 and 0.02, judging lower metallicities would be irrelevant for a nearby and evolved galaxy such as UGC~8179. For nebular emission, we used \texttt{nebular} module, where emission lines are computed using the photoionization model \texttt{CLOUDY} \citep{Ferland17,VillaVelez21}. Concerning dust attenuation, we adopted the attenuation law described by \citet{CharlotFall00}: the light of young stars ($< 10$~Myrs) suffers from the attenuation by both the dust in their birth cloud, and the ISM dust. We needed to finely sample a range in $A_V$ to (i) fit the UV-optical part of the spectrum, (ii) explore the degeneracy between the dust reddening and the light from the old stellar population, and (iii) create a wider lever arm on IR emission. Finally, the dust emission was computed from templates of \citet{Dale14} based on nearby star-forming galaxies. We adopted the fiducial value of $\alpha=2$ for the power-law slope and allowed some freedom on radiation field and PAH to fit the mid-IR measurements of the WISE bands. The intergalactic medium absorption follows the prescription of \citet{Meiksin06}. The entry parameters and module names can be found in Table~\ref{tab:param_cigale}. 

\begin{table*}[h!]
\caption{Summary of parameter values used in modules for SED fitting.}
\label{tab:param_cigale}
\centering
\begin{tabular}{llcll}
\hline\hline
Property & Parameter & Unit & Values & Description \\
\texttt{module} & & & \\
\hline
SFH                   & \texttt{tau\_main} & Gyr & (1, 2, 3, 4, 5, 7.5, 10) & \small{e-folding time of the main stellar population model}\\
\texttt{sfhdelayedbq} & \texttt{age\_main} & Gyr & 12.7 & \small{Age of the main stellar population in the galaxy}\\
                      & \texttt{age\_bq}   & Myr & (10, 25, 50, 75, 100, & \small{Age of the burst/quench episode} \\
                      & & & 300, 500)\\
                      & \texttt{r\_sfr}    & --  & (0, 0.05, 0.1, 0.5, 0.8, & \small{Ratio of the SFR after/before age\_bq}\\
                      & & & 1, 2, 5, 7, 10, 25,50, \\
                      & & & 75, 100)\\

\hline
Stellar population & \texttt{imf}         & --  & 1 & \small{IMF from \citet{Chabrier03}}\\
\texttt{bc03}      & \texttt{metallicity} & --  & (0.008, 0.02) & \small{stellar metallicity (0.4 solar or solar)}\\

\hline
Nebular emission  & \texttt{logU}         & -- & -2.0& \small{Ionization parameter}\\
\texttt{nebular}  & \texttt{zgas}         & -- & 0.02 & \small{gas metallicity (solar)}\\
                  & \texttt{ne}           & cm$^{-3}$ & 100 & \small{Electron density}\\  
                  & \texttt{f\_esc}       & -- & 0 & \small{Fraction of Lyman continuum escaping the galaxy}\\
                  & \texttt{f\_dust}      & -- & 0 & \small{Fraction of Lyman continuum absorbed by dust}\\
                  & \texttt{lines\_width} & km.s$^{-1}$ & 300 \\
                  
\hline
Dust attenuation                 & \texttt{Av\_ISM}    & mag & [0.05, 1.5] \small{in 15 steps} & \small{V-band attenuation in the ISM}\\
\texttt{dustatt\_modified\_CF00} & \texttt{mu}         & --  & (0.2, 0.3, 0.4) & \small{Av\_ISM / (Av\_BC+Av\_ISM)}\\
                                 & \texttt{slope\_ISM} & --  & -0.7 & \small{Slope of the attenuation in the ISM} \\
                                 & \texttt{slope\_BC}  & --  & -1.3 & \small{Slope of the attenuation in the birth clouds}\\
                                 
\hline
Dust emission   & \texttt{qpah}  & -- & (2.5, 3.90, 5.26) & \small{Mass fraction of PAH} \\
\texttt{dl2014} & \texttt{umin}  & Habing & (0.1, 1.0, 5.0, 10.0, &\small{Minimum radiation field}\\
                & & & 50.0) \\
                & \texttt{alpha} & -- & 2.0 & \small{Powerlaw slope dU/dM $\propto$ U$^\alpha$}\\
                & \texttt{gamma} & -- & 0.1 & \small{Fraction illuminated from Umin to Umax}\\
\hline
\end{tabular}
\end{table*}

With the grid of models (priors) computed, the module \texttt{pdf\_analysis} estimates the physical properties and the uncertainties from the likelihood-weighted means and standard-deviations.
The output properties of interest for this study are the stellar mass (\texttt{stellar.m\_star}) and SFR averaged over the past 10 and 100~Myrs (\texttt{sfh.sfr10Myrs} and \texttt{sfh.sfr100Myrs}, respectively $\SFRten$ and $\SFR100$). We preferentially rely on $\SFR100$ in the upcoming Section~\ref{sec:results}, as the photometric measurements we provided as constraints for the fit are informing us on SF on long timescales rather than ongoing SF. We further explored the influence of Balmer lines H$\alpha$ and H$\beta$ measurements from IFU-data MaNGA survey as additional constraints on the SED fitting in Sec.~\ref{subsec:discussion:SFR}. As these recombination lines are emitted by ionized gas around newly born stars, they subsequently give us insight into ongoing SF, typically measured by $\SFRten$. We discuss these results in Section~\ref{subsec:discussion:SFR}.

\subsection{Resolved SED fitting} 
\label{subsec:conv_rg}
The resolved fits of the SEDs were constrained with flux measurements in ten bands: GALEX-\textit{FUV, -NUV}, SDSS in \textit{u, g, r, i}, and \textit{z} bands, and WISE bands 1 (3.4~µm), 2 (4.6~µm), and 3 (11.4~µm). The multiwavelength photometric data described in Sec.~\ref{subsec:photo_data} come from various instruments and have therefore different angular resolution. Hence, images need to be homogenized to a common Point Spread Function (PSF) beforehand, to avoid any color aberration in the spectra. In order to address this issue, we convolved all data to the coarsest resolution with the corresponding kernel from \citet{Aniano11}\footnote{Astronomical Convolution-kernel repository:    \url{https://www.astro.princeton.edu/~draine/Kernels/Kernels_2018/Kernel_FITS_Files/Hi_Resolution/}}. The WISE4 band has the lowest spatial resolution with PSF FWHM of 12$\arcsec$ \citep{Wright10}. Convolving all the images to the resolution of WISE4 would significantly degrade the rest of the data. We thus set the WISE4 band aside as far as the resolved SED fitting is concerned and used the convolution kernels to the 6.5$\arcsec$ PSF of WISE3 instead. 

The ten background-subtracted and convolved maps (GALEX-\textit{FUV, -NUV}, SDSS-\textit{u, -g, -r, -i, -z}, WISE1, 2 and 3) were then reprojected following Nyquist sampling criterion, on a common grid of $3\arcsec\times3\arcsec$ pixels using a flux-conservative method. At $z=0.052$, $3\arcsec$ $\sim3$~kpc along the major axis, however, UGC~8179 is quite inclined ($i=65$°); hence, the deprojected surface of a pixel $\simeq 25$~kpc$^2$. The inclination angle was computed through the axis ratio. We used NED isophotal values in the $r$ band for $a$ and $b$ yielding a ratio of $b/a = 0.46$. We also assumed a typical disk thickness of $q = c/a = 0.2$, where $c$ is the intrinsic thickness of the disk. The inclination is then computed following $\sin i = \sqrt{(1-(b/a)^2)/(1-q^2)}$. Given that the source presents a position angle of 16°, both sides of the pixel suffer from projection along the line of sight. Once corrected, the physical area of the pixel is $\mathcal{A}= 24.4$~kpc$^2$.

\section{Results}
\label{sec:results}

\subsection{CO maps}
\label{subsec:CO}
\begin{figure}[!htbp]
   \centering
   \begin{overpic}[width=0.98\hsize]{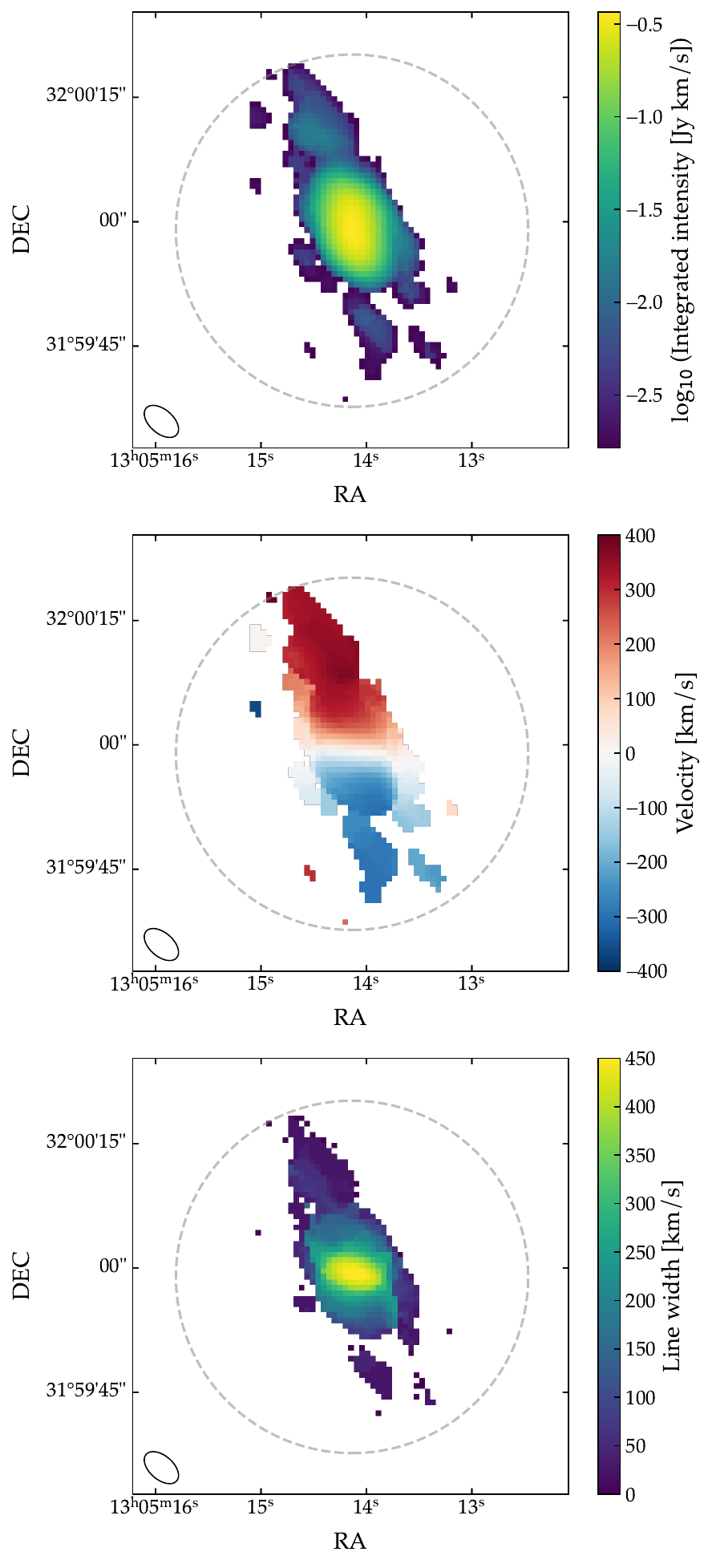}
   \put(20.3,79.4){\includegraphics[width=0.215\textwidth]{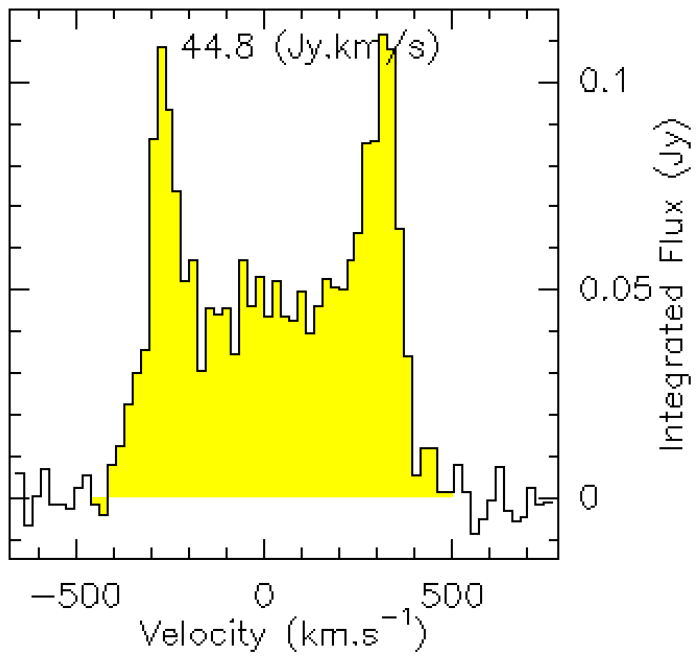}}
   \end{overpic}
    \caption{CO(1-0) moment maps from the cleaned data cube. \textit{Top}: Moment-0 in log$_\text{10}$ (integrated line intensity). Insert: Integrated spectrum at 20 km/s resolution over the region where CO reaches significant S/N. \textit{Middle}: Moment-1 ($V_\t{los}$, intensity-weighted velocity). \textit{Bottom}: Moment-2 (line width). The dashed gray circle represents the NOEMA HPBW at the redshifted CO(1-0) frequency. The  ellipse at the bottom right corner is the synthesized beam ($4.9\arcsec\times2.9\arcsec$).
   }
    \label{fig:COmomentmaps}
\end{figure}

In Fig.~\ref{fig:COmomentmaps} are shown the moment 0, 1 and 2 maps of CO detection, with a clipping at 5$\sigma$. As previously mentioned, the primary beam correction increases the noise at the edge of the FoV, we therefore applied an elliptical mask around the main detection for visualization clarity. 
In the bottom right corner of each panel is the resolution element that mainly depends on the $uv$-coverage and the exposure time. In the top right corner of the moment 0 map (top panel) we show the integrated CO line along with the integrated flux density of $S_\t{CO}\Delta V= 44.8 \pm 0.54~\t{Jy~\kms}$. As a comparison, L23 measured $49.5 \pm 1.45~\t{Jy~\kms}$ from their single-dish detection. We computed the estimation of molecular gas mass $\MH2= (1.02 \pm 0.03) \times 10^{10}$~\Msun, from our $S_\t{CO}\Delta V$ using Eq.~\ref{eq:COline} and with Eq.~\ref{eq:COtoH2} together with a metallicity-dependent conversion factor $\aCO^\text{MaNGA}$ as described in Sec.~\ref{subsec:molgas}. This result is lower than what L23 found ($\MH2 = 4.47 \pm 1.36 \times 10^{10}$~\Msun), 
but still consistent since L23 values are extrapolated molecular gas mass, assuming an exponentially decreasing profile from their single-dish detection.

The CO line is exhibiting a clear double-horn profile, typical signature of what is expected from a rotating disk \citep[e.g.,][]{Young&Scoville91,Salome15}. The integrated intensity is strongly enhanced at the center of the galaxy, and is decaying toward the outskirts of the disk. The disk appears slightly asymmetrical with less smooth high-S/N detection in the southern part of the CO-disk, yielding more clump-like features. In the moment 1 map (middle panel), the velocity profile corresponds to the line profile and confirms the rotating state of the compact detected disk. The velocity field suggests the existence of a bar, since the kinematic minor axis is not perpendicular to the major axis, which was unclear from the optical images. The velocity curve, corrected for inclination such that $V_\t{rot}=V_\t{los} / \sin(i)$, flattens at $380~\t{\kms}$ for distances from the center of 10-20 kpc -- a remarkably high value. For comparison, \citet{DiTeodoro23} recover the rotational velocity of the atomic gas $V_\t{flat} = 318 \pm 25~\t{\kms}$ through HI observations at a 15--85 kpc distance from the center. This $\sim 60~\t{\kms}$ gap between the molecular and atomic gas phases is likely tracing the transition between the baryon-dominated, high velocity mid-plane inner disk (CO) and the mildly slower, dark-matter-dominated outer disk (HI). Besides, the H$\alpha$ rotation curve measured in MaNGA, peaks at a comparably high $V_\t{rot} \sim 370~\t{\kms}$, before decreasing toward the HI lower plateau, up to a radius of 25~kpc, bridging the two regimes. 

We recover the line width in the bottom panel showing the moment 2 map, and the corresponding enhanced feature at the center of the disk corresponding to high velocity gradients. These line widths are a combination of rotational broadening and dispersion. We recover $\sigma \t{(10-20 \:kpc}) \sim 44~\t{\kms}$ estimated from the moment-2 map away from the central region, in order to limit velocity broadening contamination. We find $V_\t{rot}/\sigma=8.48$, which is notably lower than the typical values of 20-–30 reported in the outer molecular disks of nearby spirals \citep[e.g.,][]{Mogotsi16}. One should however note that our dispersion estimate -- based on moment-2 maps -- can overestimate the true turbulent dispersion in regions with significant velocity gradients or low S/N. 

\begin{figure}[h!]
   \centering
   \includegraphics[width=\hsize]{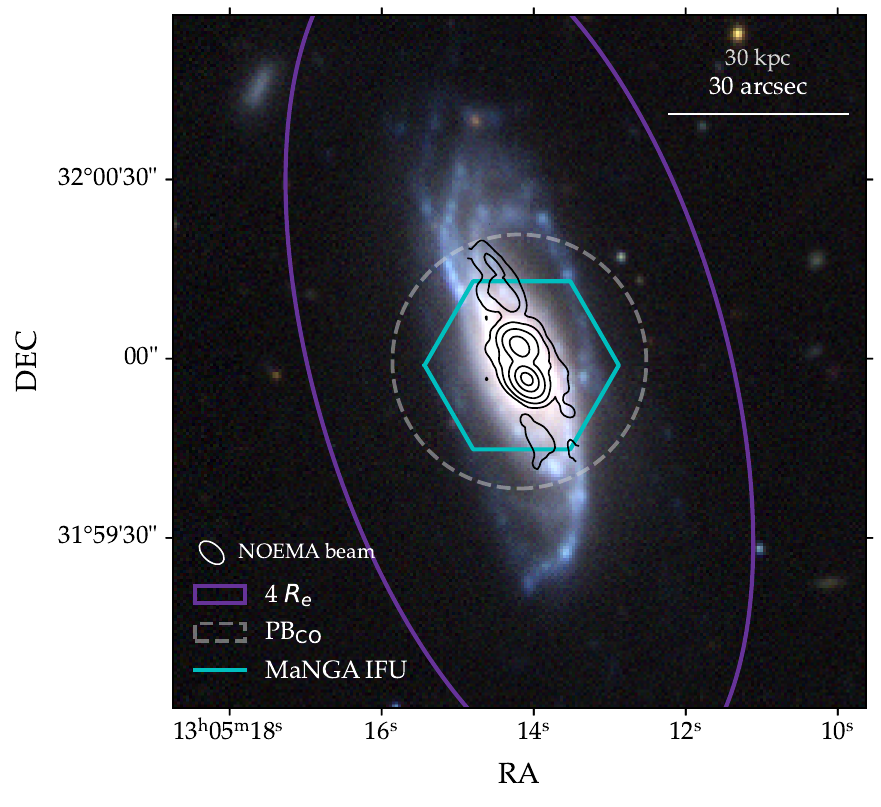}
    \caption{Distribution of CO(1-0) maxima from NOEMA data on a background optical image from the DESI Legacy Survey (DR9). The contour levels are [4.5, 9, 20, 40, 60]~mJy~beam$^{-1}$. The dashed gray circle represents the NOEMA HPBW size at the CO(1-0) redshifted frequency, while the purple ellipse indicates the semimajor axis of 4~$R_e$. The white ellipse at the bottom left corner is the synthesized beam of NOEMA ($4.9\arcsec\times2.9\arcsec$). }
    \label{fig:CO+SDSS}
\end{figure}

We look at how the molecular gas is distributed in the disk of the galaxy in Fig. \ref{fig:CO+SDSS} by overlaying contours of CO(1-0) maxima on top of the optical ($g$, $r$, and $z$ in blue, green, and red, respectively) image from the DESI Legacy Survey.
We recover the cold gas reservoir double-peaked profile at the center of the galaxy. Our detection spans the bulge and reaches the base of the spiral arms. One can note that the FoV of NOEMA (dashed gray circle; see also Fig.~\ref{fig:COmomentmaps}) is limiting the detection mainly to the inner part of the disk. Due to the disk inclination, it also encompasses some of the outer disk in the minor axis direction; however, most of the spiral arms are out of the FoV. Judging from the blue color of the arms in the optical image, the outer disk likely hosts a younger stellar population, where we expect to find an abundant H$_2$ reservoir.  Alongside the observed data are drawn three regions. The cyan hexagon shows the coverage of MaNGA IFU. The remaining two correspond to the regions where the photometric flux was extracted for the global (purple ellipse) and local (dashed gray circle) estimations of the galaxy's properties using SED fitting (see Sec \ref{sec:methods}). The purple ellipse of semi major axis 4$R_e$ encompasses the convolved fluxes in all ten bands (from \textit{FUV} to WISE4). Results of these local and global estimates are listed in Table \ref{tab:glob_prop}, and shown in Fig.~\ref{fig:glob_prop_sSFR_fmol_tdep} in purple and green diamonds.

The CO integrated intensity map allowed us to compute a \H2 mass surface density map. In order to perform a pixel-to-pixel ratio of the molecular gas mass and the output properties of resolved SED fitting, we applied the same method of spatial convolution and regridding to the \H2-mass map. We converted it to the molecular gas mass surface density $\Sigma_\t{mol}$ by dividing the result by the deprojected area of the pixels.

\subsection{Global properties}
\label{subsec:global_prop}

\begin{figure}[!htbp]
    \centering
    \includegraphics[width=\hsize] {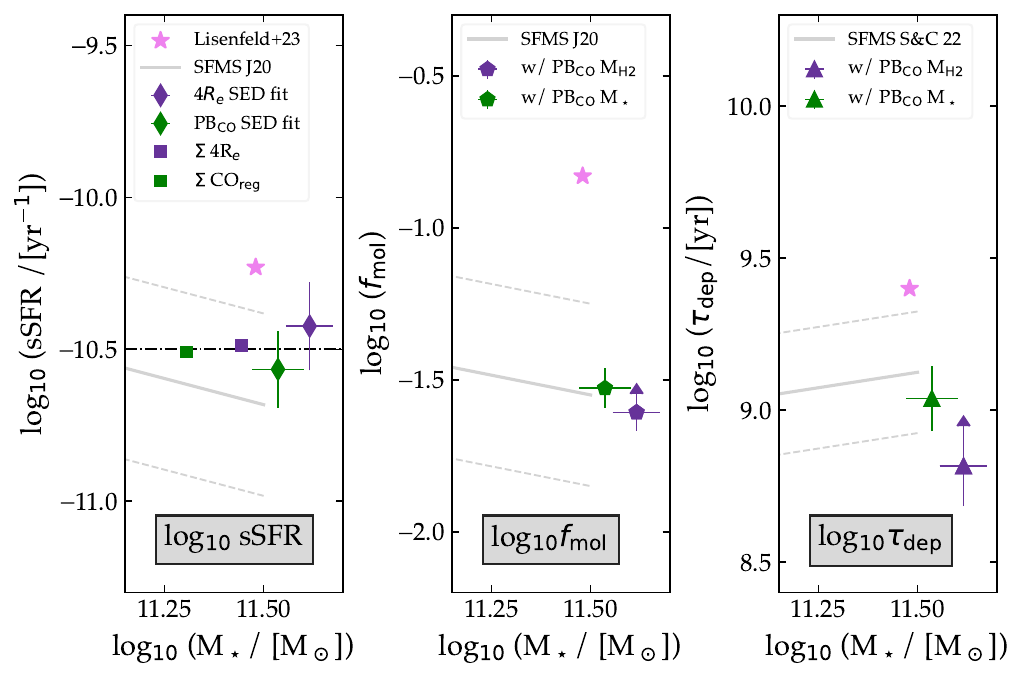}
     \caption{\textit{Left:} Log~s$\SFR100$ with respect to \M*. The green (purple) diamond and square denote the integrated fit and the sum of the resolved fits in the PB$_\t{CO}$ (4~$R_e$) region, respectively. The SFMS from \citet{Janowiecki20} is included for comparison. The dash-dotted black line shows a constant log~s$\SFR100=-10.5$ and serves as a visual guide. \textit{Middle:} Molecular gas mass fraction with respect to \M*. The purple pentagon represents the large-scale lower limit of $\fmol$. The SFMS is also from \citet{Janowiecki20}. The green pentagon corresponds to the $\fmol$ in the PB$_\t{CO}$ region only. \textit{Right:} Molecular gas depletion time with respect to \M*. The solid gray line is the linear fit for SMFS galaxies from \citet{Saintonge&Catinella22}, with a 0.2 dex scatter. The rest of the legend is identical to the middle panel, with triangles instead of pentagons. }
      \label{fig:glob_prop_sSFR_fmol_tdep}
\end{figure}

All fluxes integrated in the two regions drawn in Fig.~\ref{fig:CO+SDSS}, namely the ellipse of semimajor axis 4$R_e$, and the FoV of NOEMA, can be found in Table~\ref{tab:photo_flux}, respectively named $F_{4Re}$ and $F_\t{PBCO}$. These fluxes were the photometric measurement inputs for the best fits shown in App.~\ref{app:fits_globaux}, of which the output parameters of interest are listed in Table~\ref{tab:glob_prop}. We find $\SFR100 = 15.58 \pm 4.57~\t{\Msun~yr}^{-1}$ and $\log_{10}$~\M*/\Msun$ = 11.62 \pm 0.06$ for the entire galaxy. For comparison, L23 found a SFR = 17.8~\Msun~yr$^{-1}$ and $\log_{10}$~\M*/\Msun$ = 11.48$ following the photometric prescriptions of \citet{Janowiecki17} and \citet{Leroy19}, respectively. In the central part of the disk (PB$_\t{CO}$), we find a $\SFR100 = 9.34 \pm 2.26$~\Msun~yr~$^{-1}$ and $\log_{10}$~\M*/\Msun$ = 11.54 \pm 0.06$. In this region, we recover only a fraction of the total stellar mass, as one should expect. We also recover a lower $\SFR100$, implying that according to the photometry, there is SF occurring in the disk, where CO emission is not observed. This result yields similar specific SFR (s$\SFR100 = \SFR100$/\M*) for the two regions, as is shown in the left-hand panel of Fig.~\ref{fig:glob_prop_sSFR_fmol_tdep}. The integrated s$\SFR100$ and \M* values confirm that the spiral UGC~8179, despite its very high mass, forms stars at a rate allowing it to stand on the SFMS \citep[from][]{Janowiecki20}. Both our values and those from L23 place UGC~8179 on the more star-forming side of the SFMS, confirming that (i) it is star-forming, despite what we expect from the great majority of galaxies at this mass range (i.e., early-type galaxies), and (ii) its sSFR follows the relation of \citet{Janowiecki20}, even though it is inferred from lower-mass star-forming spirals. This suggests no particular change in scaling relations for spiral galaxies in this mass-range. Moreover, the self-consistency of the SED fitting method is confirmed by the good agreement between the global fits and the sum of the resolved fit outputs in the same area regarding sSFR. 

Together with our $\MH2$ estimate, this central \M* value allowed us to compute the local $\fmol$ in the region where CO is detected: $\log_{10}\fmol^\t{PBCO} = -1.53 \pm 0.07$. Unfortunately, the outer part of the disk is not encompassed by NOEMA's FoV. Therefore, we inferred a lower limit of $\fmol$ at the scale of the entire galaxy by taking the ratio of the central $\MH2$ with the global \M* estimate in 4~$R_e$. We find $\log_{10}\fmol \geq -1.61 \pm 0.06 $. This bounding value appears as a purple pentagon in the middle panel of Fig.~\ref{fig:glob_prop_sSFR_fmol_tdep}, along with the value found by L23 ($\log_{10}\fmol^\t{L23} = -0.83$) inferred from the exponentially extrapolated H$_2$-mass from their unresolved CO-detection. The latter value is exceeding our lower limit. It is important to remember that although they use a very conservative constant conversion factor of $\aCO=3~\t{\Msun} (\t{K~kms~pc}^{2})^{-1}$, we adopted a metallicity-dependent $\aCO$ that is systematically lower to their value in our detection region (see the right-hand panel of Fig.~\ref{fig:app:variation_KS_aCO}), yielding by definition a lower $\fmol$. Also, we measured a lower integrated line intensity than their extrapolated value, i.e., we do not recover the molecular gas mass expected from their extrapolation. Thus even with the same $\aCO$, one would obtain a lower $\fmol$. This result places UGC~8179 in a less extreme regime, as the $\fmol$ values are located within the scatter of the SFMS, where one would expect a galaxy of such mass, based on the relation inferred from SFMS galaxies and from model predictions \citep[see also][]{Saintonge17}. 

A similar computation can be carried out with the $\SFR100$ value to compute $\tdep$. This quantity is inversely proportional to the star-forming efficiency (SFE) $\epsilon=\t{SFR}\times t_{ff}/\MH2=t_{ff}$/$\tdep$, i.e., the amount of the SF per unit molecular gas (normalized by the free-fall time, which rules the physics of gas collapse in dense environments where SF takes place). We find a local $\log_{10}\tdep^\t{PBCO} = 9.04 \pm 0.11$ (i. e. $1.09 \pm 0.26$~Gyr), and lower bound of $\log_{10}\tdep \geq 8.82 \pm 0.13 $, which is consistent with that shown in \citet{Saintonge&Catinella22}.
Our values are notably lower than the mean value L23 found for their sample of 46 super spirals: $\log_{10}\tau_\t{dep}^\t{L23}= 9.30 \pm 0.03$; and don't agree with their conclusion, namely that SSGs show an overall lower SFE than their comparison sample of SFMS galaxies \citep[xCOLDGASS,][]{Saintonge11a,Saintonge11b,Saintonge17}. On the contrary, we recover a central $\tdep^\t{PBCO}$ typical of SFMS galaxies. 
These values appear as triangles in Fig.~\ref{fig:glob_prop_sSFR_fmol_tdep} and are summarized in Table \ref{tab:glob_prop}.

\begin{table}[h!]
\caption{Properties computed by SED fitting.}
\label{tab:glob_prop}
\centering
{\renewcommand{\arraystretch}{1.3}
\begin{tabular}{lcccc}
\hline\hline
         & SFR$_\t{100}$ & SFR$_\t{10}$ & $\log_{10}$M$_\star$ & $\chi^2_\t{red}$\\
         & {\small M$_\odot$ yr$^{-1}$} & {\small M$_\odot$ yr$^{-1}$} & {\small $\log_{10}$(M$_\odot$)} & \\
\hline
4 $R_e$     & 15.58 ± 4.57 & 10.29 ± 6.26 & 11.62 ± 0.06 & 0.15\\
PB$_\t{CO}$ & 9.34 ± 2.26 & 4.76 ± 3.52 & 11.54 ± 0.06 & 0.24\\
\hline\hline

        & $\log_{10}\tdep^{100}$ & $\log_{10}\tdep^{10}$ & $\log_{10}\fmol$\\
\hline
4 $R_e$     & {\small $\geq 8.82 \pm 0.13$}  &  {\small$\geq 9.00 \pm 0.31$}  &  {\small$\geq-1.61 \pm 0.06 $} \\
PB$_\t{CO}$ & $9.04 \pm 0.11$ & $9.33 \pm 0.41$ & $-1.53 \pm 0.07$\\
\hline
\end{tabular}
}
\tablefoot{These results were obtained after flux extraction on the convolved maps (see Section~\ref{subsec:conv_rg}) in an ellipse of semimajor axis $a=4~R_e$ and in the region of the primary beam at NOEMA's HPBW (see Fig. \ref{fig:CO+SDSS}).}
\end{table}

Thus, when comparing the three panels of Fig. \ref{fig:glob_prop_sSFR_fmol_tdep}, we notice that the regular fraction of molecular gas available to fuel the formation of new stars is likely balanced by the rather low molecular gas depletion time. This results in UGC~8179 standing in the more active side of the SFMS.

\subsection{Resolved properties}
\label{subsec:resolved_prop}

\begin{figure*}[h!]
    \centering
    \includegraphics[width=\hsize] {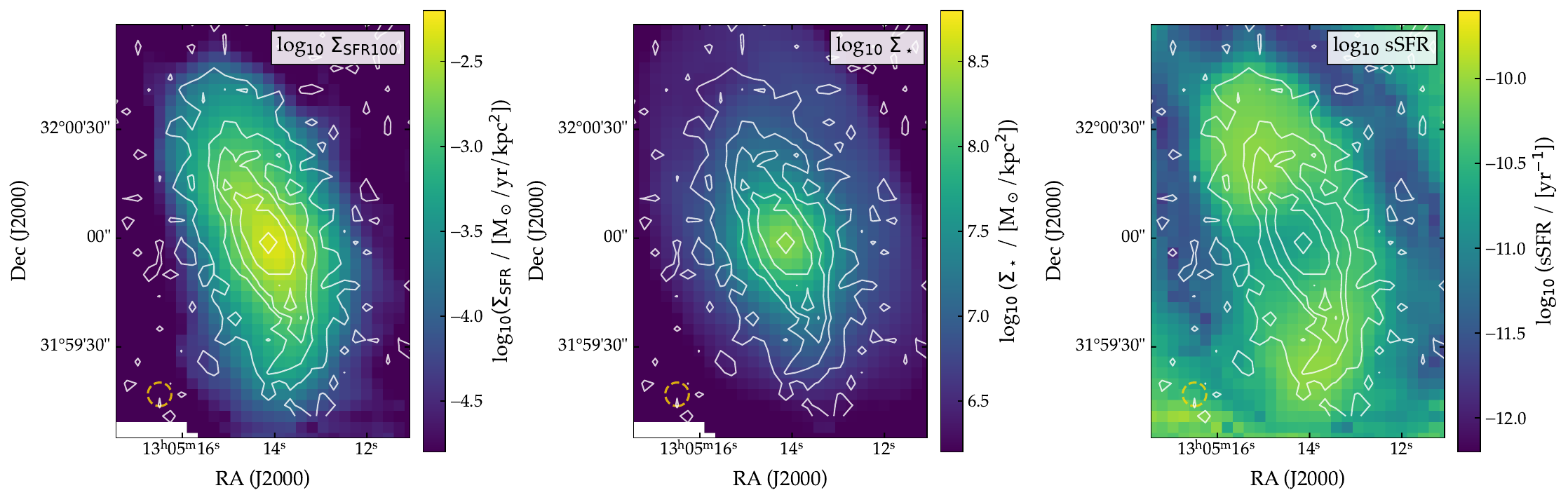}
     \caption{Mapped results of the resolved SED fitting run at a 3$\arcsec$ resolution. The white contours show the emission distribution in the SDSS $g$ band. The dashed yellow circle represents the PSF at the FWHM of $WISE3$. \textit{Left}: Log surface density of the Bayesian estimate of the SFR averaged over the past 100 Myrs. \textit{Center}: Log surface density of the Bayesian estimate of M$_\star$. \textit{Right}: Specific $\SFR100$. }
      \label{fig:SFR_Mstar_sSFR}
\end{figure*}

The mapped surface densities of SFR$_{100}$ and \M* are shown in Fig. \ref{fig:SFR_Mstar_sSFR}, along with the ratio of both quantities, i.e., the sSFR$_{100}$.
Both $\Sigma_\t{SFR100}$ and $\Sigma_\star$ exhibit a peaked feature at the center of the galactic disk, although $\Sigma_\star$ decays more steeply with radius, resulting in enhanced regions of sSFR$_{100}$ in the spiral arms and a shallow dip in the center. We recover what is described in resolved studies of spiral galaxies: sSFR drops in the inner regions of the disk and even more so for high-mass spirals \citep[$11 < \log_{10}\t{M*/\Msun}< 12$,][]{Abdurrouf17,Belfiore18,Coenda19}, exhibiting a decrease of 0.5 to 1~dex. This central dip likely reflects an ongoing suppression of SFR in the nucleus, in line with morphological-quenching scenarios. 

\begin{figure}[!htbp]
   \centering
   \includegraphics[width=0.7\hsize]{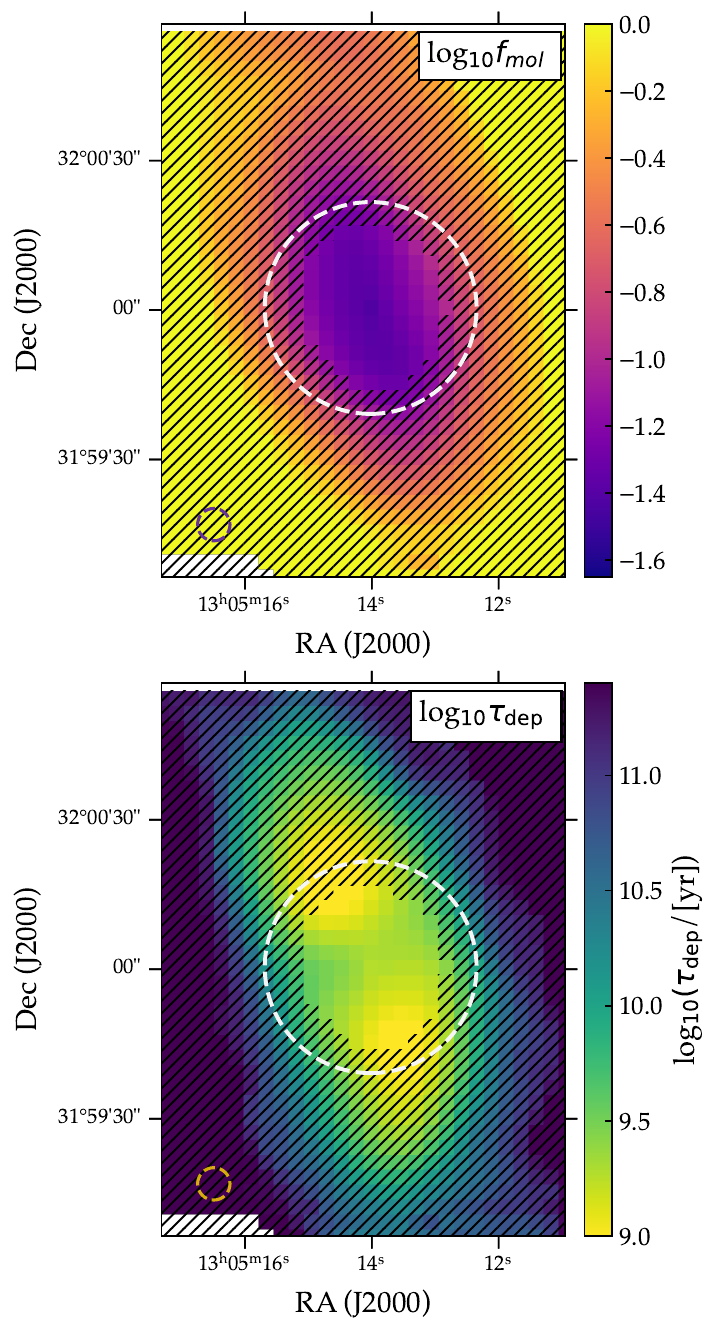}
    \caption{\textit{Top}: Log $\fmol$, the molecular gas mass fraction. \textit{Bottom}: Log $\tau_\text{dep}$, the molecular depletion time. The hatched pixels are below a detection threshold of 4$\sigma$, where $\sigma$ is the noise in the CO(1-0) moment 0 map. The dashed white circle represents NOEMA's the HPBW.}
    \label{fig:fmol_tdep}
\end{figure}

As previously done for the global $\fmol$ and $\tdep$, the resolved $\fmol$ and $\tdep$ are obtained through a pixel-to-pixel ratio of the resolved map of $\MH2$ with the maps of \M* and SFR$_{100}$, respectively. The resulting maps are shown in Fig.~\ref{fig:fmol_tdep}.
The hatched pixels correspond to pixels below the 4$\sigma$ detection threshold in the CO-integrated intensity map. The molecular gas fraction exhibits a declining profile toward the center and reaches $\log_{10}\fmol\sim-1.4$. This value is in agreement with the lower bound to the global $\fmol$ mentioned above. On the other hand, $\tdep$ reveals an saddle-like profile. Indeed, an increase appears in the inner part of the detection region, whereas lower $\tdep$ regions are hinted in the spiral arm region. Assuming the inside-out growth scenario, we would expect SF to be suppressed in the bulge region and be more active in the spiral arms -- although it is not what \citet{Querejeta21} show in their resolved study of $\tdep$ in SFMS galaxies. 

\subsubsection{Radial profiles}
\begin{figure*}[h!]
    \centering
    \includegraphics[width=\hsize] {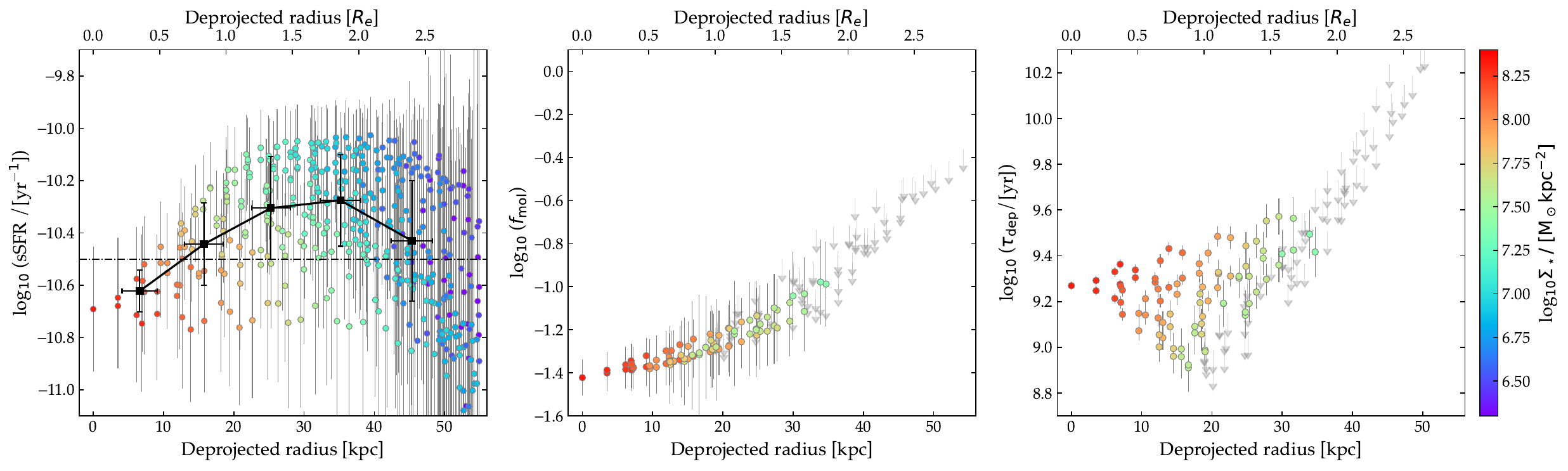}
     \caption{Radial profiles. \textit{Left:} Log$_{10}$ s$\SFR100$.  \textit{Middle}: Log$_{10}\fmol$. \textit{Right}: Log$_{10}\tdep$. The gray arrows correspond to the upper limits computed for the hatched pixels within PB$_\t{CO}$.}
      \label{fig:radial_profiles_sSFR_fmol_tdep}
\end{figure*}

\paragraph{Specific SFR.} In the left panel of Fig. \ref{fig:radial_profiles_sSFR_fmol_tdep} we show the results from the resolved SED fits mapped in the right panel of Fig.~\ref{fig:SFR_Mstar_sSFR}, here plotted as circles, color-coded with stellar mass density.
 The black squares show the uncertainty-weighted means in 10 kpc radial bins. Here, the centrally declined sSFR mentioned in Sec.~\ref{subsec:resolved_prop} appears more clearly. It is similar to the median radial profile found by \citet{Pan24} for ALMaQUEST star-forming galaxies (classified as SF$_\t{h}$ and SF$_\t{m}$). UGC~8179 corresponds to the SF$_\t{h}$ category with a global $\log_{10}~\t{sSFR} > -10.5$, for which the corresponding sSFR median radial profile is very similar to that of UGC~8179, both in magnitude and downturn amplitude. Moreover, it is in line with the 0.5-1~dex drop discussed by \citet{Belfiore18}, more specifically for high-mass SFMS galaxies (see Fig.~3 and 7 therein, 11 < $\log_{10}$~\M* < 11.5). 
The large-scale s$\SFR100$ plotted as the dash-dotted black line at log$_{10}$s$\SFR100=-10.5$, is similar to the local s$\SFR100$ found in the inner ($\lesssim$ 0.8 $R_e$) disk, where the SF might be suppressed compared to the rest of the disk. 
\vspace{-4mm}
\noindent \paragraph{Molecular gas mass fraction.} In the middle panel of Fig. \ref{fig:radial_profiles_sSFR_fmol_tdep}, we show the radial profile of $\fmol$.
One can notice the decrease in $\fmol$ of 0.5~dex toward the center of the disk, across 30 kpc (i.e., $\sim 1.5~R_e$) and a decade of \M*. This profile indicates that the molecular gas density and the stellar mass density profile have different scale lengths, and could hint at a higher stellar density in the center of the disk. 
On the other hand, when considering the upper limits (gray arrows), the radial increase is more likely an artifact of dividing a constant noise value for $\MH2$ by a decreasing profile for stellar mass in the outskirts of the PB$_\t{CO}$ region. 
In any case, these values should be considered for what they are: mere upper limits.
\vspace{-4mm}
\noindent\paragraph{Molecular gas depletion time.} The right panel of Fig.~\ref{fig:radial_profiles_sSFR_fmol_tdep} shows the radial profile of $\tdep$. The resolved values reveal a different profile than the monotonous slope of $\fmol$. Indeed $\tdep$ exhibits an overall decreasing profile, with however a hint of increase toward the center of the galaxy. This could also be due to greater scatter rather than a change of slope, which a saddle-like feature could induce (Fig.~\ref{fig:fmol_tdep}). The apparent bend occurs at $r\sim 20~\t{kpc} \sim 1~R_e$, and it is not clear whether there is a corresponding $\Sigma_\star$ at which the $\tdep$ changes behavior. As previously mentioned, such an increase in $\tdep$ is not expected from the previous results in the literature \citep[e.g.,][]{Leroy08,Querejeta21}. 
Depletion time radial variations, including with different $\aCO$ calibrations are further discussed in Sec.~\ref{subsec:discussion:alphaCO}. 

\subsubsection{Scaling relations}
\label{subsec:scaling_relation}

\begin{figure*}[!htbp]
    \centering
    \includegraphics[width=\hsize] {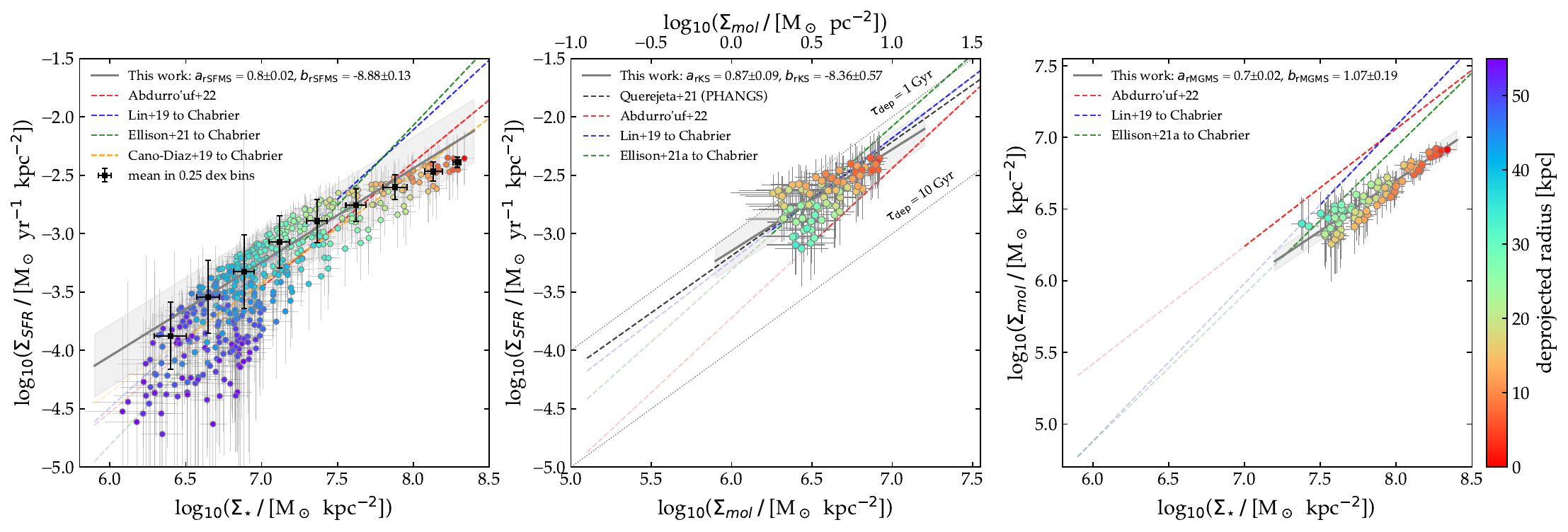}
     \caption{\textit{Left:} Resolved SFMS. 
     Each circle point represents a pixel from Fig. \ref{fig:SFR_Mstar_sSFR} with a cutoff at radii greater than 55 kpc, color-coded as a function of distance to the galactic center. The black squares represent the uncertainty-weighted mean data points in the 0.25 dex \M* bins, the error bars show the scatter within the mass bin, and the thick solid gray line is the uncertainty-weighted (ODR) linear fit to the data. The shaded area traces the 1$\sigma$ scatter around the linear fits. We included linear relations of resolved SFMS from the literature in dashed colored lines (\citet{Abdurrouf22} in red, \citet{Lin19} in blue, \citet{CanoDiaz19} in yellow, and \citet{Ellison21} in green). 
     \textit{Center:} Resolved Kennicutt-Schmidt relation. The gray line represents the ODR linear fit of the data points above the 4$\sigma$ detection threshold. The dashed black line indicates the fit to the PHANGS sample across all morphological masks from \citet{Querejeta21}. 
     \textit{Right:} Resolved MGMS. The legend is similar to the middle panel. }
      \label{fig:rSFMS_rKS_rMGMS}
\end{figure*}

We combined the spatially resolved estimates of $\SFR100$ and \M* with the molecular gas mass measurements to study the relationships between the SF, molecular gas mass, and stellar mass on kiloparsec scales. In Figure \ref{fig:rSFMS_rKS_rMGMS}, we compile the three resolved relations for UGC~8179: the resolved SFMS (rSFMS, i.e., $\Sigma_\t{SFR}$ -- $\Sigma_\star$, left panel), the resolved molecular Kennicutt-Schmidt (rKS, i.e., $\Sigma_\t{SFR}$ -- $\Sigma_\t{mol}$, middle panel) and the resolved Molecular Gas Main Sequence (rMGMS, i.e; $\Sigma_\t{mol}$ -- $\Sigma_\star$, right panel).
Studying these relations in a resolved manner, namely at the scales where SF happens, gives enlightening insight into the local physical processes that drive and regulate SF. The colored circles and black squares have the same legend as in Fig.~\ref{fig:radial_profiles_sSFR_fmol_tdep}.

We fit our data using the orthogonal distance regression (ODR) fitting method to take into account data uncertainties on both axes, with the following power laws:
\begin{align}
    &\log_{10}\Sigma_\t{SFR100} = a_\t{rSFMS} \times \log_{10}\Sigma_\star + b_\t{rSFMS},\\
    &\log_{10}\Sigma_\t{SFR100} = a_\t{rKS} \times \log_{10}\Sigma_\t{mol} + b_\t{rKS},\\
    &\log_{10}\Sigma_\t{mol} = a_\t{rMGMS} \times \log_{10}\Sigma_\star + b_\t{rMFMS}.
\end{align}
We report the computed values in Table~\ref{tab:param_scaling_relations} along with the values from the comparison studies. We assume a Chabrier IMF, whereas some of the featured results estimates SFRs from H$\alpha$ measurements assuming a Salpeter IMF. Ionizing photons are mostly produced by \M* $ > 10~\t{\Msun}$ stars; hence, H$\alpha$-related SFRs estimates are particularly sensitive to the IMF. Thus, an offset was applied both to SFR and \M* values as follows: SFR$_\t{C03}$ = SFR$_\t{S55}\times 0.63$ and \M*$_\t{,C03}$ = \M*$_\t{,S55} \times 0.61$ \citep{Madau&Dickinson14}.
\vspace{-4mm}
\paragraph{Resolved SFMS.} The linear fit of the rSFMS, shown as the solid gray line in the left-hand panel of Fig.~\ref{fig:rSFMS_rKS_rMGMS}, reveals no particular offset from the other studies it is compared to. Its slope is however notably lower: we find $a_\t{rSFMS}=0.80\pm0.02$. 
\citet{CanoDiaz19}, \citet{Ellison21}  and \citet{Lin19} (from now on referred as C19, E21 and L19) find $a_\t{rSFMS} \sim [0.94 ~, 1.19]$, and rely on H$_\alpha$ luminosities for SFR estimates. \cite{Abdurrouf22} (A22 hereafter) use a method similar to ours, inferring properties from pixel-to-pixel SED fitting; however, their spectral coverage extends from UV-light to far-infrared wavelengths. All pixels radially follow the trend of the SFMS, with higher-mass pixels located in the center of the disk and forming stars at a higher rate per unit surface. Our data exhibit a bend in the slope around $\log_{10}\t{M}_\star \sim 7.2$, and although the outer part of the disk ($30 \lesssim r \t{(kpc)}\lesssim 55$) shows a steeper slope resembling other studies' results, our fit is more tightly constrained by the inner radii part ($r \lesssim 30$ kpc), where the sSFRs decrease. Part of the galaxies in E21 exhibit such a bend (see Fig. 3 therein).

In parallel to this rSFMS single linear fit, we explored the option of two separate scaling relations. The parameters of the two ODR fits are referred to as $a_\t{rSFMS}^{lm}$ and $a_\t{rSFMS}^{hm}$ for the low- and high-mass regimes, respectively, and the results are plotted in the left panel of Figure~\ref{fig:rSFMS_double_fit}. Both of the curves are encompassed by the initial single fit scatter. We find that the low-mass fit is in better agreement with the literature with a slope of $a_\t{rSFMS}^{lm} = 1.14 \pm 0.04$, compared to the single linear fit. The decrease in sSFR and the subsequent change of slopes from 1.14 to 0.59 is discussed in Sec.~\ref{subsec:disc:central_sSFR}.

\vspace{-4mm}
\paragraph{Resolved KS.} We examined the correlation between the SFR the molecular gas mass surface density, known as the molecular Kennicutt-Schmidt relation \citep{Kennicutt98b,Schmidt59}. It has been extensively studied in various objects and at diverse scales in galaxies to probe how molecular gas can drive SF. Its slope is well constrained, nearly unity \citep[e.g.,][]{Bigiel08,Leroy13}, yielding a universal molecular $\tdep$ of $\sim$ 2 Gyr.  
The rKS of UGC~8179 is shown in the middle panel of Fig. \ref{fig:rSFMS_rKS_rMGMS}, along with a linear fit as the solid gray line. The data points are solely the pixels above the detection threshold in $\MH2$. For comparison, we added the resolved fits from some of the studies aforementioned along with that of \citet{Querejeta21} (Q21 hereinafter) for MS galaxies of the PHANGS-ALMA sample. The slope of the rKS is sensitive to the resolution of the data \citep{Calzetti12,Pessa21}. More precisely, above a spatial resolution threshold of $\sim 1-2~\t{kpc}$, the relation tends to neighbor unity. Thus, we rigorously chose to compare our result with kiloparsec-scale resolved studies. A22, E21, and L19 resolve kiloparsec scales, whereas we attain $\sim 6~\t{kpc}$. However, we remain in the same regime as the comparison studies, where star-forming regions are unresolved and properties are spatially averaged. We thereby consider the small resolution gap to have negligible impact on the scaling relations. The slope we find is lower than unity: $a_\t{rKS}=0.87\pm0.09$, yet consistent with the literature. This galaxy shows an overall comparable molecular depletion time to that of the lower-mass MS spirals, except for a few lower stellar mass-density pixels that lean toward greater $\tdep$ ($\sim 3-4$~Gyr). 
\vspace{-4mm}
\paragraph{Resolved MGMS.} We observe in the right-hand panel of Fig.~\ref{fig:rSFMS_rKS_rMGMS} the tight correlation between the local molecular gas and stellar surface densities. It has the least scatter ($\sigma_\t{rMGMS} = 0.07$~dex), followed by the rKS and lastly the rSFMS with $\sigma_\t{rKS} = 0.15$~dex and $\sigma_\t{rSFMS} = 0.27$ dex, respectively. Our data points are overall consistent with the literature; however, our value for the slope is remarkably lower than unity: $a_\t{rMGMS}=0.70 \pm 0.02$, which implies a varying $\fmol$ with \M* -- a radius as pointed out above. Similarly, A22 find a slope lower than unity ($a_\t{rMGMS}^\t{A22}=0.82 \pm 0.01$), whereas E21 and L19 slopes are linear to superlinear, without taking into account near-IR light in their \M* estimate. 

Following the reasoning of a double scaling law in the rSFMS, namely in $\Sigma_\t{SFR}$ with respect $\Sigma_\star$, we assumed that the local SF processes are universal in the galaxy, i.e., direct proportionality, and applied our rKS relation parameters to both high and low-mass regimes to extrapolate $\Sigma_\t{mol}$ from the measured $\Sigma_\t{SFR}$ \citep[a similar method is used by][]{Bigiel08}. From that, a double-slope rMGMS emerges, displayed in the right panel of Fig.~\ref{fig:rSFMS_double_fit} as thick dashed lines. The predicted fit for the low-stellar mass regime, where we did not observe molecular gas, lies in very good correspondence with the continuity of E21 and L19 fits ($a_\t{rMGMS}^{lm} = 1.09$).
The mass at which the apparent transition in SFR occurs is $\log_{10}(\Sigma_\star / \t{[\Msun~kpc}^{-2}]) \simeq 7.2 \pm 0.1$, which corresponds to a  $\log_{10}(\Sigma_\t{mol} / \t{[\Msun~kpc}^{-2}]) \simeq 6 $, i.e., $\Sigma_\t{mol} \simeq 1$ \Msun~pc$^{-2}$. 

The slopes of the linear rSFMS and rMGMS relations we derived for UGC~8179 show a clear negative offset compared to the relations derived for nearby star-forming spirals at the same kpc-resolution. Thus, the lower $\fmol$ shown in the rMGMS could reflect the slightly less-replenished molecular gas reservoir, especially toward the center, yielding lower SFR values in the rSFMS in this region, despite similar SFE. As the linear fit is more tightly constrained by the smaller uncertainties in this central region, this results in the flattening of the rSFMS. This is compatible with the conclusion of \citet{Lin19}, namely that the rSFMS might emerge from the combination of the rKS and the rMGMS. In any case, the rKS is in good agreement with the values of the same studies, implying that the underlying physical processes of SF and its efficiency are consistent.

\begin{table}[h!]
\caption{Resolved scaling relation slope, intercept, and scatter values for the rSFMS, rKS, and rMGMS, for this and other studies.}
\label{tab:param_scaling_relations}
\centering

\begin{tabular}{lccc}
\hline\hline
  Reference & Slope & Intercept & Scatter\\
\hline 
& $a_\t{rSFMS}$ & $b_\t{rSFMS}$ & $\sigma_\t{rSFMS}$ \\
\hline
UGC~8179 & $0.80 \pm 0.02$ &  $-8.88 \pm 0.13$ & 0.27 \\
A22 & $1.07 \pm 0.02$ & $-10.95 \pm 0.17$ & 0.31\\
E21\tablefootmark{a} & $1.37 \pm 0.01$ & $-13.03 \pm 0.1$ & 0.28 \\
L19\tablefootmark{a} & $1.19 \pm 0.01$ & $-11.63 \pm 0.11$ & 0.25 \\
CD19\tablefootmark{a} & $0.94 \pm 0.08 $& $-10.0 \pm 0.69$ & 0.27\\
\hline
UGC~8179 $lm$ & $1.14 \pm 0.04$ & $-11.21 \pm 0.3$ & 0.28 \\
UGC~8179 $hm$ & $0.59 \pm 0.05$ & $-7.29 \pm 0.23$ & 0.16 \\
\hline
& $a_\t{rKS}$ & $b_\t{rKS}$ & $\sigma_\t{rKS}$ \\
\hline
UGC~8179 & $0.87 \pm 0.09$ & $-8.36 \pm 0.57$ & 0.15\\
A22     &  $1.28 \pm 0.01$ & $-11.40 \pm 0.09$ & 0.19\\ 
E21\tablefootmark{a} & $1.23 \pm 0.01$ & $-10.69 \pm 0.06 $ & 0.22 \\
L19\tablefootmark{a} & $1.05 \pm 0.01$ & $-9.53 \pm 0.06$ & 0.19 \\
Q21 & $0.976 \pm 0.053$ & $-3.189 \pm 0.009$ & 0.24-0.35 \\
\hline
& $a_\t{rMGMS}$ & $b_\t{rMGMS}$ & $\sigma_\t{rMGMS}$ \\
\hline
UGC~8179 & $0.70 \pm 0.02$ & $-1.07 \pm 0.19$ & 0.07\\
A22 & $0.82 \pm 0.01$ & $ 0.50 \pm 0.07$ & 0.19\\
E21\tablefootmark{a} & $1.03 \pm 0.01$ & $-1.30 \pm 0.06$ & 0.21\\
L19\tablefootmark{a} & $1.1 \pm 0.01$ & $-1.72 \pm 0.08$ & 0.20 \\
\hline
\end{tabular}
\tablefoot{
    \tablefoottext{a}{Intercepts values were corrected to the Chabrier IMF when the parent study assumes a Salpeter IMF.}}
\end{table}

\section{Discussion}
\label{sec:Discussion}

\subsection{Influence of conversion factor $\alpha_\text{CO}$}
\label{subsec:discussion:alphaCO}

Molecular gas mass estimation has a major uncertainty: the conversion factor from CO luminosity to \H2 mass, namely $\aCO$ \citep{schinnerer2024}. This subsequently affects the molecular gas depletion time $\tdep$ as well as the molecular gas fraction $\fmol$. 
In the Milky Way, \citet{Bolatto13} recommend as a constant value $\alpha_\text{CO}^\text{MW}=4.3$~M$_\odot$(K~km.s$^{-1}$~pc$^{2}$)$^{-1}$. We applied this value to our analysis to compare with our results. 

We also included the conservative value of $\alpha^\t{L23}_\text{CO}$ = 3~$M_\odot$~(K~km~s$^{-1}$~pc$^{2}$)$^{-1}$ from L23 in the comparison. 

As shown in Figure~\ref{fig:metallicity_radprofile_MaNGA}, the metallicity is supersolar over the FoV of NOEMA, reaching a $12 + \log(\t{O/H}) \simeq 9 \simeq 2~Z_\odot$ in the innermost region. This metallicity is the upper bound to the validity range of the -1.5 index in eq.~\ref{eq:alphaMaNGA} \citep{Saintonge&Catinella22}. This results in very low $\aCO$ values across our molecular gas detection region ($< 2~R_e$), $\aCO^\t{MaNGA}$ spans $\sim 1.4-2.9$ \Msun~(K~km~s$^{-1}$~pc$^{2}$)$^{-1}$ (see right-hand panel of Fig.~\ref{fig:app:variation_KS_aCO}). Thus, adopting a constant $\aCO$ yields an overestimation of the mass of molecular gas, even when choosing the low value used by L23.

The rKS linear fits with the constant $\aCO^\t{MW}$ and $\aCO^\t{L23}$ are shown in the top left panel of Fig.~\ref{fig:app:variation_KS_aCO} to compare with our results from $\aCO^\t{MaNGA}$. The values of the slopes and intercepts of the ODR linear fits are listed in Table \ref{tab:discussion:KS_slopes_alphaCO}. 
\begin{table}[h!]
    \caption{Resolved Kennicutt-Schmidt relation with varying $\aCO$.}
    \label{tab:discussion:KS_slopes_alphaCO}
    \centering
    {\renewcommand{\arraystretch}{1.5}
    \begin{tabular}{llcc}
    \hline\hline
      & & Slope $a_\t{rKS}$ & Intercept $b_\t{rKS}$\\
    \hline
    $Z$-dependent & $\alpha_\text{CO}^\text{MaNGA}(r)$ & $0.87 \pm 0.09$ & $-8.36 \pm 0.57$ \\ 
    \hline 
    Constant & $\aCO^\t{MW} = 4.3$ & $0.70 \pm 0.05$ & $-7.49 \pm 0.37$   \\
             & $\aCO^\t{L23} = 3$ & $0.70 \pm 0.05$ &  $-7.38 \pm 0.37$ \\
    \hline
    \end{tabular}
    }
\end{table}

When comparing the L23 and MW $\aCO$ results, one can easily see that a lower $\aCO$ implies a smaller $\MH2$, thus a shorter $\tdep$ to meet the considered SFR. Furthermore, these constant $\aCO$ are associated with slopes that are less steep for the corresponding rKS fits, since the central pixels have greater molecular gas mass. This results in longer $\tdep$ than with a metallicity-dependent $\aCO$, and induces a varying $\tdep$ with $\Sigma_\t{mol}$, i.e., lower SFE for high molecular gas surface density. However, \citet{Bigiel11} showed that $\tdep$ is overall constant with $\Sigma_\t{mol}$, i.e., that the underlying SF processes are uniform in galactic discs.  

\begin{figure}[!htbp]
    \centering

    \includegraphics[width=\hsize]{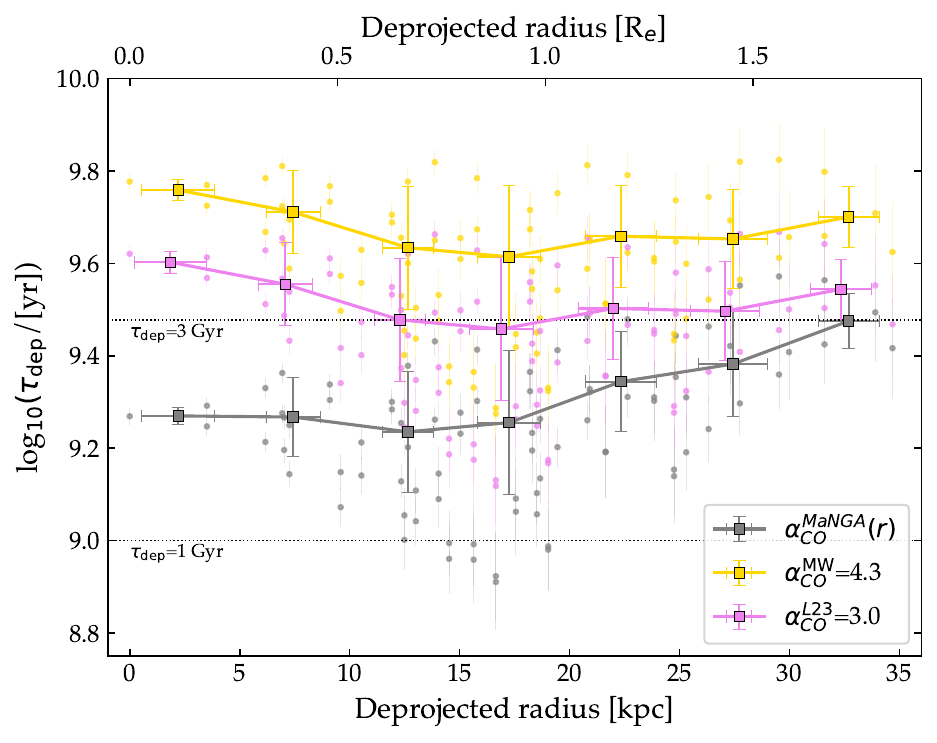}
    \caption{Radial profiles of molecular gas depletion time $\tdep$ with varying $\aCO$. The resolved measurements are shown as circles, and the uncertainty-weighted means in 10 kpc radius-bins (from [0-5] to [30-35] kpc) are shown as squares. The error bars of the squares correspond to the data scatter in each direction within the radius bin considered. A slight negative offset in the radius of the means was added to the pink curve for clarity.}
    \label{fig:discussion:radial_tdep}
\end{figure}

We further explore the spatial distribution of $\tdep^{100}$ by tracing the radial profiles for all $\aCO$ prescriptions in Figure \ref{fig:discussion:radial_tdep}.
The radial $\tdep$ trend with the metallicity-dependent $\aCO$ shows a decreasing profile toward the center, namely what most studies find in SFMS spiral galaxies. For example, Q21 find a slight drop in $\tdep$ within the centers of PHANGS-ALMA galaxies, and a median $\tdep = 1.18 \pm 0.65$~Gyr, compared to $\tdep = 1.79^{+1.04}_{-0.74}$~Gyr in the spiral arms, observed as a $\sim$ 0.25~dex vertical offset in the two corresponding local rKS relations. Other studies on nearby star-forming spirals agree on a decreasing molecular $\tdep$ toward the center/bulge region, where the gas is often molecular and where the stellar potential is larger \citep{HuangKauffmann15,Utomo17}.
On the other hand, \citet{Longmore13} argue that in the Galactic center ($r \lesssim 500$~pc), where the surface density of cold gas reaches a maximum, SF is slowed down, supposedly by an additional turbulent energy that prevents the gravitational collapse of the gas. 
On the other hand, when considering constant $\aCO$, regardless of its value, the combination of the radial profiles of SFR and $\MH2$ yields an increasing $\tdep$ trend toward the center of the disk, as suggested by the low slope values of the rKS (see Table \ref{tab:discussion:KS_slopes_alphaCO}). Although a constant $\tdep$ radial profile seems in reasonable agreement with our data, given the large error bars corresponding to the scatter in the radial bins (5~kpc), the linear ODR fits for both constant $\aCO$ (not shown here) result in a decreasing trend with increasing radius. We note that these results exclude the possibility of an increasing trend with radius, such as what was found with the metallicity-dependent $\aCO$, and as revealed by most studies in SFMS spiral galaxies.

Our metallicity measurements come with their own uncertainties, nevertheless, we show how taking currently available information and calibrations into account would impact our calculations and conclusions. We find that it is essential to adopt a conversion factor that takes metallicity into account for UGC~8179 considering its unusually high metallicity. A constant $\aCO$ would lead to spurious conclusions on the SF conditions and processes in the disk, and most importantly in the central region.

\subsection{Lower sSFR in the center}
\label{subsec:disc:central_sSFR}

As suggested by \citet{Martig09}, SFE could be lowered in a galaxy with a large bulge-to-disk ratio, thanks to the stabilization of the gas against collapse by the bulge's gravitational potential (i.e., the morphological quenching scenario). Several studies have shown the link between bulge-growth and the onset of SF inefficiency at fixed stellar mass, implying that although a bulge is not sufficient, it is a required condition to explain quenching \citep{DiMauro22,Eales20,Lang14}. Therefore, we decomposed the light profile of UGC~8179 in two components (bulge and disk) using the 2D morphology fitting code \texttt{galfit} \citep{Peng02,PengCY10} in order to estimate a bulge-to-total luminosity ratio $B/T$ as described in \citet{Freundlich19} (Section~3.4 therein). They carried out the fits on HST ACS images in the F814W $I$ band for galaxies at $z = 0.5 - 0.8$, which correspond to rest-frame SDSS-$r$ and -$g$ bands. We applied their method on the SDSS-$r$ band background-subtracted, galactic extinction-corrected images. We set the sersic indices of the disk and the bulge to be 1 and 4, respectively. For UGC~8179, we found a bulge-to-total ratio of $B/T=0.43 \pm 0.05$, typical of intermediate spirals (Sb, Sbc). Thus, this bulge could physically cause the downturn of sSFR in the central 1~$R_e$. It also is compatible with the inside-out growth paradigm, stating an earlier formation in the central region, later on reaching higher radii, as the center becomes more evolved and shows lower sSFR \citep[e.g.,][]{deJong96,Lilly&Carollo16,Goddard17}. \citet{Abdurrouf17} argue similarly that the central decline of sSFR is consistent with an older population of stars in the bugle. It is worth noting, however, that despite the drop in sSFR our results do not indicate that the SFE is affected by the bulge, as $\tdep$ keeps decreasing toward the center (Fig.~\ref{fig:discussion:radial_tdep}).

The residuals between the model and the observations reveal the spiral arms along with an oblong shape at the center of the disk, which supports our assumption of a stellar bar (mentioned in Sec.~\ref{subsec:CO}). \citet{Gavazzi15} argue that the presence of a strong bar plays an important role in rapidly quenching the SF in the central kpc of discs, for massive galaxies above a $z$-dependent threshold \M* $> \t{M}_\t{knee} \propto (1+z)^2 (= 10^{9.5}$ for local galaxies). Such a stellar structure can be responsible for turbulence and shear and thereby delay SF. Moreover, we do not rule out that potential past starburst episodes might have induced strong stellar feedback in the bulge region, subsequently heating and dispersing the molecular gas. 

On the other hand, the presence of a bulge implies higher stellar density in the inner disk. This can drag the inner most pixels toward higher $\Sigma_\star$ in the rSFMS and the rMGMS, thereby flattening the relations. This increase in bulge-mass has already been invoked in past studies to explain the sublinearity of the SFMS at high \M* and thus could partly explain the observed bend in our rSFMS, as well as our lower slope in the rMGMS. 

\subsection{Uncertainties in SFR}
\label{subsec:discussion:SFR}
\begin{SCfigure*}[1][h]
\centering
    \includegraphics[width=1.45\hsize] {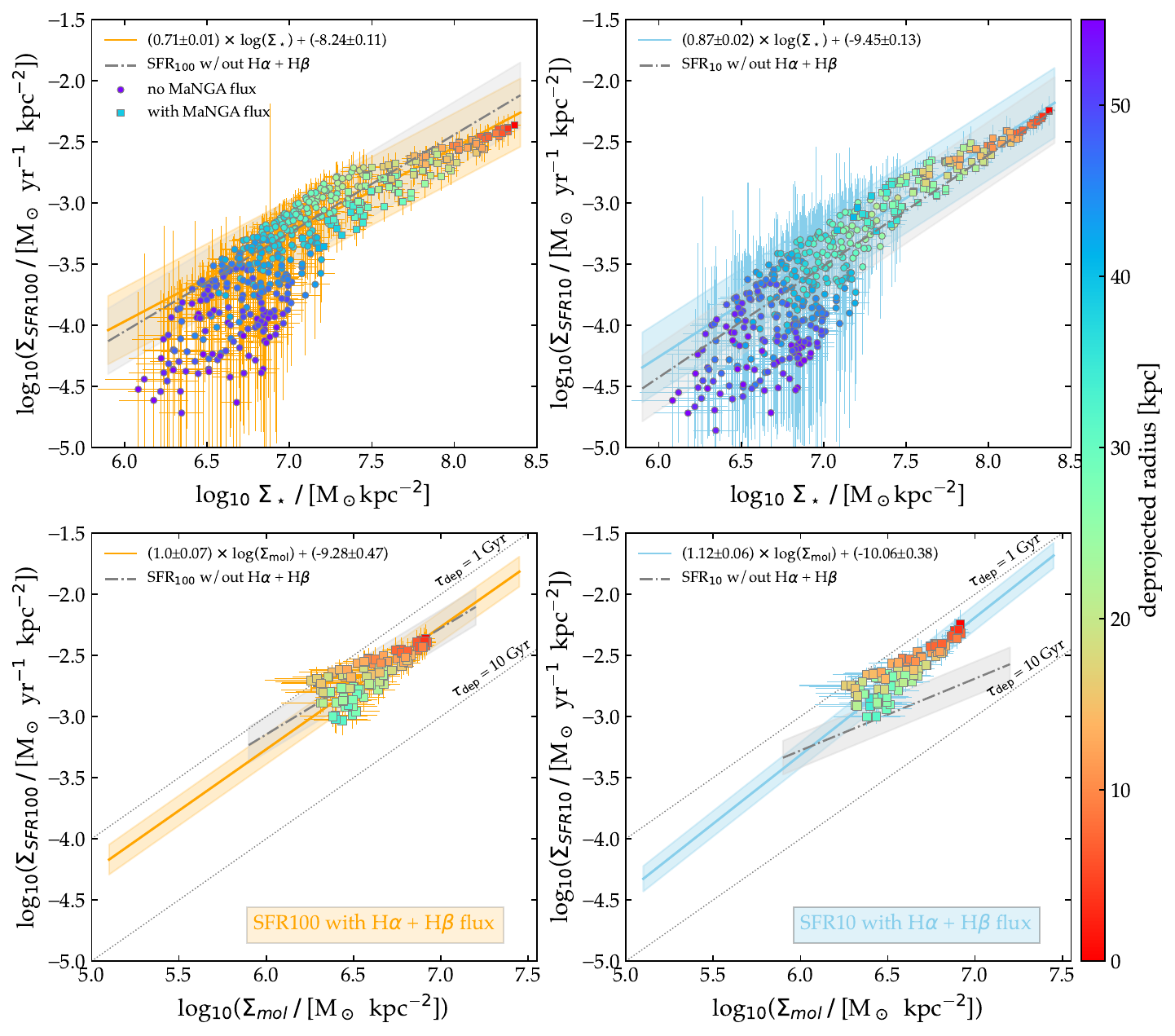}
     \caption{\textit{Top:} Resolved SFMS. The dash-dotted gray line in the left panel shows the fiducial ODR fit for $\Sigma_\t{SFR100}$ without adding the H$\alpha$ and H$\beta$ fluxes (left-hand panel of Fig.~\ref{fig:rSFMS_rKS_rMGMS}). The dash-dotted gray line in the right panel likewise indicates the fit for $\Sigma_\t{SFR10}$ without the additional flux. The squares denote the pixels affected by the addition of the MaNGA fluxes; the circles are unchanged (outside the FoV). The thick solid orange and blue lines and the shaded area respectively represent the ODR linear fits and the 1$\sigma$ scatter for the squares and circles.\\
     \textit{Bottom:} Resolved Kennicutt-Schmidt relation. The squares and circle are the same as above. The dash-dotted gray line in both panels shows the ODR fit for $\Sigma_\t{SFR100}$ (left) and $\Sigma_\t{SFR10}$ (right) without adding the H$\alpha$ and H$\beta$ fluxes.} 
      \label{fig:discussion:rSFMS_rKS_SFRprior}
\end{SCfigure*}

Except for A22 and Q21, the comparison studies in this paper use H$_\alpha$ luminosity as a proxy for SFR. There are limitations to the estimation of SFRs through H$_\alpha$ luminosity, as SF is not the only source of ionizing photons. Indeed, other excitation mechanisms such as AGNs or diffuse ionized gas (DIG) can contribute to the H$_\alpha$ luminosity, thereby leading one to overestimate the SFR. Using the Balmer decrement would also lead to overcorrection, making it appear that H$\beta$ is more attenuated than it truly is. In the BPT diagram of UGC~8179 (see Fig.~\ref{fig:BPT_MaNGA}), we do not see a clear indication for AGNs, only light mixing toward the LINER regime, yet not in the central region. Additionally, no radio continuum was detected in the NOEMA data, weakening even more the potential H$\alpha$ contamination by the AGNs in the case of this galaxy. 

However, in this method, the calibration constant relies on the assumption of a continuous SF period of at least 100~Myr \citep{Kramarenko25}, while H$\alpha$ only traces young stars ($< 10$~Myrs) and SFHs can vary and the calibration constant with it \citep{Boquien14}. Another key assumption concerns the ionizing photons absorption by dust, namely that the latter absorbs no Lyman continuum photons. Yet, studies show that ionizing photons are indeed absorbed by dust and helium, leaving $\sim 57$\% to ionize the hydrogen \citep{Tacchella22}. This hypothesis thus leads one to overestimate the SFR in dusty environments, and MaNGA measurements of UGC~8179 indicate such a dust-rich bulge. However, we exploited the wealth of IFS observations in our study.

We investigated the robustness of our SFR estimate from SED-fitting by modeling new SEDs while adding H$\alpha$ and H$\beta$ fluxes for the pixels concerned. The MaNGA survey (Mapping Nearby Galaxies at Apache Point Observatory) is an integral-field spectroscopic survey of $\sim$10\,000 nearby galaxies \citep{bundy2015}. It uses the 2.5 m Sloan telescope \citep{gunn2006} and the BOSS spectrographs \citep{smee2013} with a spectral coverage of $3\,600\,\mathrm{\textup{\AA}}$ to $10\,300\,\mathrm{\textup{\AA}}$. UGC~8179 is part of the ``color enhanced'' main sample of MaNGA, designed to populate the less dense parts of the NUV-i versus $M_i$ color-magnitude diagram. The observations are further described in table \ref{tab:MaNGA_overview}. Although the IFS has a more constraining FoV than NOEMA's HPBW, its $\sim 32\arcsec$ extent (once convolved and reprojected to the 3$\arcsec$-wide pixel grid) covers all the pixels where the CO signal exceeds the $4\sigma$ detection threshold. The H$\alpha$ and H$\beta$ fluxes were added on a separate run of CIGALE, allowing us to compare the results with and without the ionized gas measurements. This comparison is illustrated in Figure \ref{fig:discussion:rSFMS_rKS_SFRprior}, where the colored linear fits and points are the new outputs of CIGALE, which include MaNGA line measurements, and the dash-dotted gray lines are the original fiducial outputs, without the additional constraints. The impact of these new fluxes on both $\SFRten$ and $\SFR100$ are shown in Fig. \ref{fig:discussion:rSFMS_rKS_SFRprior}, in the left and right column, respectively, for the two relevant scaling relations, namely rSFMS (top row) and the rKS (bottom row).

Focusing on the rSFMS, we notice that the uncertainties dramatically diminish for the pixels with new flux inputs, and even more so for $\SFRten$, ensuring us that the latter is not a reliable quantity without a proxy of recent and ongoing SF such as H$\alpha$ and H$\beta$ fluxes. This is further confirmed by the significant raise in $\SFRten$ after the addition of H$\alpha$ and H$\beta$ fluxes, easily noticeable in the rKS ($\Delta a^{10}_\t{rKS} = a^{10\t{,H}\alpha \t{H}\beta}_\t{rKS} -a^{10}_\t{rKS} = 0.53$). This allows to recover a linear fit closer to unity, namely $a_\t{rKS}^{\t{10,H}\alpha \t{H}\beta} = 1.12 \pm 0.06$ (compared to $a_\t{rKS}^\t{10}=0.59 \pm 0.08$ without the added fluxes). Therefore, it appears CIGALE misses some of the recent SF (i.e., in the past 10 Myrs) without the ionized gas fluxes entries.

On the other hand, we find no significant difference in $\SFR100$ and \M* when adding the H$\alpha$ and H$\beta$ fluxes, as the offsets of the linear fits are enclosed within the scatter of our fiducial results. We note that the rKS with $\SFR100$ (bottom-left panel) exhibits a slightly steeper slope with MaNGA measurements than without ($a_\t{rKS}^{\t{100,H}\alpha \t{H}\beta} = 1.00 \pm 0.07$). This reduces the slight gap with comparison studies, which base their SFR on H$\alpha$ measurements: $a_\t{rKS}^\t{L19} = 1.05$, and $a_\t{rKS}^\t{E21} = 1.23$.

Lastly when comparing $\SFR100$ to $\SFRten$, we remark the larger error bars in $\SFRten$ on the circles, namely the pixels not concerned by the ionized gas inputs, which illustrates why we relied on $\SFR100$ throughout this study. Additionally, the scaling relations exhibit differences when considering the two timescales. The bend in the rSFMS is less obvious when looking at $\SFRten$. Moreover, the rKS exhibits a slightly steeper slope when considering recent SFR (with $\SFRten$) rather than $\SFR100$.

In conclusion, the addition of ionized gas data to the flux inputs of the SED fits is critical when it comes to recent SFR (namely $\SFRten$), but $\SFR100$ and \M* estimates are robust without it. It is worth keeping in mind the timescale on which the SFR is measured when comparing resolved scaling relations, as we recover different slopes with $\SFRten \t{ and } \SFR100$, i.e., $\Delta a_\t{rSFMS}^\t{10-100}=0.11$ and $\Delta a_\t{rKS}^\t{10-100}=-0.28$ without MaNGA entries, and $\Delta a_\t{rSFMS}^{\t{10-100, H}\alpha \t{H}\beta}=0.15$ and $\Delta a_\t{rKS}^{\t{10-100, H}\alpha \t{H}\beta}=0.12$ with additional H$\alpha$ and H$\beta$ fluxes.

\section{Summary and conclusions}
\label{sec:conclusions}
\citet{Ogle16} estimate that SSGs represent 6\% of the galaxies in their mass range at $z<0.3$, thereby constituting a rare population of objects whose evolutionary pathways differed from classical galaxy evolution scenarios. Studying SSGs offers substantial insights into the mechanisms that normally shut down SF in massive galaxies. In the course of this study, we investigated the resolved molecular gas content of UGC~8179, a nearby ($z=0.052$) super massive, actively star-forming spiral galaxy, along with its global and resolved star-forming properties. Our main results are:
\begin{itemize}
    \item We detected a replenished molecular gas reservoir through CO observations with the IRAM/NOEMA interferometer. It allowed us to measure a molecular gas mass of $\MH2 = (1.02 \pm 0.03) \times 10^{10}$ \Msun.
    \item We find that in the case of UGC~8179, it is essential to adopt a metallicity-dependent conversion factor, considering its very high metallicity. We showed how a constant $\aCO$, even a very conservative value such as that adopted by L23, could lead to incorrect conclusions about the SF conditions and processes in the disk -- and most importantly, in the central region.
    \item We developed a SED-fitting-based method using CIGALE to probe the global and resolved star-forming properties of massive, extended, and unquenched spiral galaxies, utilizing archival photometric data from the UV to the mid-IR wavelengths range. The spatial resolution adopted was limited by the PSF of the $WISE3$ band to $3\arcsec\times3\arcsec$ pixels, which results in a deprojected surface of $\sim~25$~kpc$^2$ per pixel. The consistency of this method from local to global scales is confirmed by the good agreement between the latter and the integrated properties of the resolved outputs.
    \item Our global $\SFR100$ and \M* estimates confirm that despite its large stellar mass ($\log(\t{\M*/\Msun})=11.62 \pm 0.06$), UGC~8179 is on the high-mass end of the SFMS with $\SFR100=15.68 \pm 4.57$ \Msun~yr$^{-1}$. In addition, we confirm its gas-rich nature with a molecular gas mass fraction of $\log_{10}\fmol \geq -1.61 \pm 0.06$, even though we recover a less extreme value than L23. Similarly, we do not reach their conclusion of a higher $\tdep$: we measure $\log_{10}\tdep^{100}\geq 8.82$ and a local estimate of $\log_{10}\tdep^{100}=9.04 \pm 0.11$ in the inner disk, consistent with MS spiral galaxy measurements.
    \item We investigated the three SF-related resolved scaling relations (namely the rSFMS, the rKS, and the rMGMS relations) for UGC~8179 and compared them to the SFMS galaxies' resolved scaling-relations from the literature. We find all quantities to be well correlated. The very specific characteristics of this SSG (its large mass and spatial extent) allowed us to explore low-surface density regimes that are difficult to access and thus little studied ($\Sigma_\star< 10^7~\t{\Msun~kpc}^{-2}$ and $\Sigma_\t{SFR100} < 10^{-3.5}~\t{\Msun~yr}^{-1}\t{~kpc}^{-2}$). 
    \item More specifically, we find a rKS relation in reasonable consistency with unity ($a_\t{rKS} = 0.87 \pm 0.09$). Our data matches the literature fits, which suggests that standard local SF processes are at play in the molecular gas of this SSG.
    \item The rSFMS is reasonably compatible with other studies at similar spatial kiloparsec resolutions, albeit at a lower slope ($a_\t{rSFMS} = 0.80 \pm 0.02$) when using a single scaling relation. This is because the rSFMS seems to experience a change in behavior in the bulge region and a corresponding decrease in the sSFR radial profile in the center. Fitting the relation with two components makes the relation even more consistent with the rSFMS from the literature up to values of $\log_{10}(\Sigma_\star/\t{[\Msun~kpc}^{-2}])< 7.2 $. Above this value, a second but shallower relation is possible and could be explained by the buildup of a non-star-forming bulge at a higher stellar surface density and smaller radial distances. 
    \item This flattening is also present in the rMGMS in the inner disk and could be partly induced by the enhanced $\Sigma_\star$ due to the bulge, as the $\fmol$ also suggests. The plausible presence of a stellar bar hints at a dynamically regulated SF, rather than a purely gas-regulated inner disk. However, the rMGMS is less constrained than the other scaling relations because of the absence of molecular gas detections at low surface densities ($\Sigma_\star < 10^7~ \t{\Msun~kpc}^{-2}$ and $\Sigma_\t{mol} <10^6~\t{\Msun~kpc}^{-2}$, a regime at which the transition from atomic gas to molecular gas occurs.)
\end{itemize}

UGC~8179 is part of a greater sample of galaxies constituted by SSGs and less massive, yet remarkably baryon-rich massive unquenched spirals. We chose to develop on this remarkably extended and star-forming target and thereby explore a low $\Sigma_\star$ range seldom investigated, even in local spirals. In a forthcoming paper, we shall apply this method to the rest of the sample (19 sources) and explore whether discrepancies arise from its morphology, environment, or nuclear activity. A comprehensive survey of the molecular gas content in a galaxy such as UGC~8179 at greater radii could significantly improve our resolved study of its properties across its extended disk.

\begin{acknowledgements}
The authors thank the referee for their constructive comments, that improved the clarity of the paper. RC acknowledges funding support from the Initiative Physique des Infinis (IPI), a research training program of the Idex SUPER led by the Alliance Sorbonne Université. SF acknowledges funding support through a scholarship of the German Academic Exchange Service (DAAD), a scholarship of the Heinrich Böll Foundation and a mobility stipend of the Paris Observatory - PSL University. PS acknowledges support by the French National Research Agency (ANR-25-CE31-5364). MB acknowledges support by the ANID BASAL project FB210003. This work was supported by the French government through the France 2030 investment plan managed by the National Research Agency (ANR), as part of the Initiative of Excellence of Université Côte d’Azur under reference No. ANR-15-IDEX-01. UL acknowledges support by the research grant PID2023-150178NB-I00, financed by MCIU/AEI/10.13039/501100011033 and from the Junta de Andaluc\'ia (Spain) grant FQM108.
This work is based on observations carried out under project number S23BP with the IRAM NOEMA Interferometer. The authors would like to thanks IRAM staff for the help provided during the observations and for data reduction. IRAM is supported by INSU/CNRS (France), MPG (Germany) and IGN (Spain).
This work was made possible by the NASA/IPAC Extragalactic Database and the NASA/ IPAC Infrared Science Archive, which are both operated by the Jet Propulsion Laboratory, California Institute of Technology, under contract with the National Aeronautics and Space Administration.
This work is based on observations made with the NASA Galaxy Evolution Explorer (GALEX), which is operated for NASA by the California Institute of Technology under NASA contract NAS5-98034.
Funding for the Sloan Digital Sky Survey V has been provided by the Alfred P. Sloan Foundation, the Heising-Simons Foundation, the National Science Foundation, and the Participating Institutions. SDSS acknowledges support and resources from the Center for High-Performance Computing at the University of Utah. SDSS telescopes are located at Apache Point Observatory, funded by the Astrophysical Research Consortium and operated by New Mexico State University, and at Las Campanas Observatory, operated by the Carnegie Institution for Science. The SDSS web site is \url{www.sdss.org}. SDSS is managed by the Astrophysical Research Consortium for the Participating Institutions of the SDSS Collaboration, including the Carnegie Institution for Science, Chilean National Time Allocation Committee (CNTAC) ratified researchers, Caltech, the Gotham Participation Group, Harvard University, Heidelberg University, The Flatiron Institute, The Johns Hopkins University, L'Ecole polytechnique f\'{e}d\'{e}rale de Lausanne (EPFL), Leibniz-Institut f\"{u}r Astrophysik Potsdam (AIP), Max-Planck-Institut f\"{u}r Astronomie (MPIA Heidelberg), Max-Planck-Institut f\"{u}r Extraterrestrische Physik (MPE), Nanjing University, National Astronomical Observatories of China (NAOC), New Mexico State University, The Ohio State University, Pennsylvania State University, Smithsonian Astrophysical Observatory, Space Telescope Science Institute (STScI), the Stellar Astrophysics Participation Group, Universidad Nacional Aut\'{o}noma de M\'{e}xico, University of Arizona, University of Colorado Boulder, University of Illinois at Urbana-Champaign, University of Toronto, University of Utah, University of Virginia, Yale University, and Yunnan University.  
This publication makes use of data products from the Wide-field Infrared Survey Explorer, which is a joint project of the University of California, Los Angeles, and the Jet Propulsion Laboratory/California Institute of Technology, and NEOWISE, which is a project of the Jet Propulsion Laboratory/California Institute of Technology. WISE and NEOWISE are funded by the National Aeronautics and Space Administration.
This research uses of the NASA/IPAC Infrared Science Archive, which is funded by the National Aeronautics and Space Administration and operated by the California Institute of Technology.

\end{acknowledgements}

\bibliographystyle{bibtex/aa} 

\bibliography{bibtex/references}

\appendix

\section{Metallicity with MaNGA}
\label{app:MaNGA_SFR}

We used the already reduced observations from the data release DR17, using version 3.1.1 of the data reduction pipeline \citep[DRP][]{law2016}. We later on used emission line products from the data analysis pipeline \citep[DAP, version 3.1.0][]{westfall2019, belfiore2019, law2021}. The quantities mentioned in this study are derived from Gaussian emission line fits.
We accessed these data products through \texttt{marvin} (version 2.8.0) and \texttt{sdss\textunderscore access} (version 3.0.4) Python packages \citep{cherinka2019, cherinka2024}. Information about UGC 8179 within SDSS-IV MaNGA is summarized in Table \ref{tab:MaNGA_overview}.

\begin{table}[h]
  \centering
    \caption{Central properties of the archival IFS observation of UGC~8179 within SDSS-IV MaNGA.}
    \label{tab:MaNGA_overview}
    \begin{tabular}{lccc}
    \hline
    plateifu & FoV$^1$ & PSF FWHM & IFU design \\
      (ID)   & ($R/R_e$)  & (kpc) & \\
    \hline
    12484-12705 & 0.99 – 2.16  & 2.89 &  \makebox[0pt][l]{\raisebox{-0.25ex}{\includegraphics[height=1.7ex]{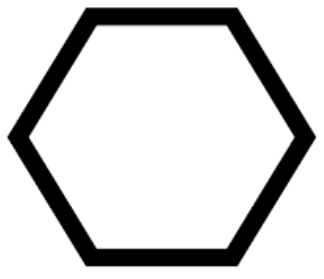}}} \hspace{1.5ex} 127-fiber\\
    \hline
    \end{tabular}
    \tablefoot{$^1$FoV defined as maximal radial coverage along the major axis $a$ (min) and minor axis $b$ (max), in units of effective radii $R_e$.}
\end{table}

To derive the metallicity 12+log(O/H) and its radial profile, we employed the strong line method with the hybrid calibration of the flux ratio $R23$ based on the direct $T_e$ method and photoionization models from \cite{maiolino2008}. 
It is double valued, but it can be safely assumed that UGC~8179 lies on the high metallicity branch:
\begin{align}
    R23 &= \log \frac{F([OII]\lambda\lambda 3726, 28) + F([OIII]\lambda\lambda 4959,5007)}{F(H\beta)} \label{eq:R23}.
\end{align}
We closely followed the approach defined in \citet{belfiore2017} to select star-forming regions and to correct fluxes for attenuation (similar to the above). It should be noted that the calibration is valid for HII regions only.
We then binned the metallicity by deprojected radius and computed the inverse-variance-weighted means. 

\begin{figure}
      \centering
      \includegraphics[width=\linewidth]{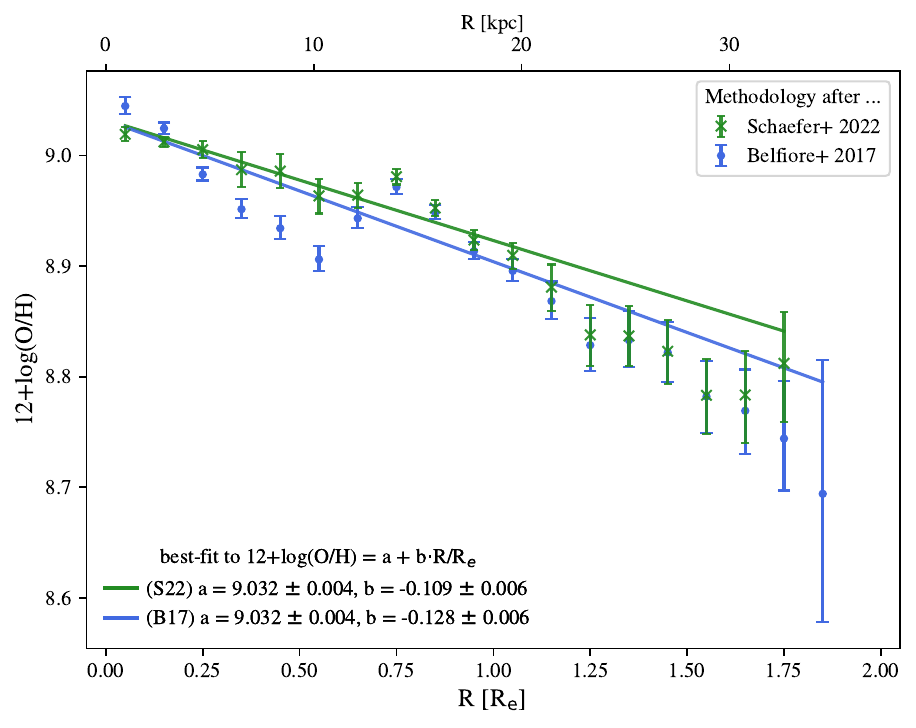}
      \caption{Metallicity as a function of deprojected radius (bottom axis) and effective radius $R_e$ (top axis, value from Table \ref{tab:IDcard})}. The median value per radial bin (width $0.1\,\mathrm{R_e}$) was computed following the methodology of \citet{schaefer2022} \citep{belfiore2017}, shown as green crosses (blue circles). The solid lines show the linear fits.
      \label{fig:metallicity_radprofile_MaNGA}
   \end{figure}

Gradients to the metallicity profiles are fitted in the radial range of 0.5 to 2$\,R_e$, due to a flattening in the center, and observational effects such as beam-smearing and inclination \citep{belfiore2017}. We obtained a best fit through an inverse-variance-weighted least-squares fit assuming Gaussian symmetric errors, while the uncertainty is characterized in a nonparametric way via residual-bootstrapping. The result is shown is Figure \ref{fig:metallicity_radprofile_MaNGA}.

The derived metallicities and their gradients are consistent with mass trends reported by \cite{belfiore2017}. 
We then derived a radially varying CO-to-H$_2$ conversion factor from it following:
\begin{align}
    \aCO^\t{MaNGA} (Z, \Sigma_*) = \alpha_{\rm CO,MW}^{(1-0)} \left(\frac{Z}{Z_\odot}\right)^{-1.5}\cdot \left(\frac{\max (\Sigma_*, 100\,\mathrm{M_\odot\,pc^{-2}})}{100\,\mathrm{M_\odot\,pc^{-2}}}\right)^{-0.25}.
\end{align}

This is based on the calibrations given in \citet{schinnerer2024}, using \citet{hu2022} and \citet{chiang2024}. 
We assume/take $Z/Z_\odot = 10^{12+\log(O/H)-8.69}$ \citep[solar value taken from][]{asplund2009}. We didn't include the second ("starburst") term in our $\aCO^\t{MaNGA}$ computation, taking into account the stellar density, since our values are neighboring 100 \Msun pc$^{-2}$ in the region considered (see middle panel of Fig. \ref{fig:SFR_Mstar_sSFR}).

As more massive galaxies tend to have formed earlier (downsizing), we do expect significant older stellar populations and DIG contamination. This would also explain the shift toward the composite or LINER-mixing region in the BPT diagrams \citep{belfiore2022}. However, also high-metallicity HII regions - as observed here - can appear as composite or AGN-like in BPT diagnostics \citep{kewley2013}. We furthermore followed the method described in \citet{schaefer2022} to derive metallicity with a more stringent selection of HII regions based on EW(H$\alpha)>10\,\mathrm{\textup{\AA}}$, minimizing DIG contamination effects, but find consistent results.

\section{Details on photometric data}
\subsection{Conversion to Jansky}
\label{app:Jyconversion}
\subsubsection{SDSS}
The native unit of SDSS files are in nanomaggies, and the conversion to jansky is straightforward: $1~\mathrm{nmgy} = 3.631 \times 10^{-6}$~Jy.

\subsubsection{GALEX}
The conversion from counts per seconds (CPS) to Jy was made first by using the conversion factor to erg.s$^{-1}$.cm$^{-2}$.Å$^{-1}$ provided by the GALEX Observer's Guide, namely:
\begin{align*} 
\text{FUV}: F \;[\text{erg.s}^{-1}.\text{cm}^{-2}.\text{\AA}^{-1}] = 1.40 \times 10^{-15} \times \text{CPS} \\
\text{NUV}: F \;[\text{erg.s}^{-1}.\text{cm}^{-2}.\text{\AA}^{-1}] = 2.06 \times 10^{-16} \times \text{CPS}
\end{align*}
Then we followed the standard relation to convert erg.s$^{-1}$.cm$^{-2}$.Å$^{-1}$ into Jy : 
\begin{equation*}
\begin{split}
    f_\nu \: [\text{Jy}] & =10^{26} \; . \; \frac{\Delta\lambda}{\Delta\nu} \; . \; f_\lambda \: [\text{W.m}^{-2}.\text{m}^{-1}] \\
    & = 10^{26} \; . \; \frac{\Delta\lambda}{\Delta\nu} \; . \; 10^{7} f_\lambda \: [\text{erg.cm}^{-2}.\text{s}^{-1}.\text{\AA}^{-1}],
\end{split}
\end{equation*}
where $\Delta\lambda$ and $\Delta\nu$ are the bandwidth measured in meter and in Hertz, respectively.

\subsubsection{WISE}
The WISE data comes in units of DN (Digital Number), and conversion to Jy was performed following the factor provided in the Table 1 of Section~IV.3.a of the Explanatory Supplement to the AllWISE Data Release Products\footnote{\url{https://wise2.ipac.caltech.edu/docs/release/allwise/expsup/}}: DN-to-Jy conversion factor [Jy/DN] $f(\mathrm{W1}) = 1.935 \times 10^{-6}$, $f(\mathrm{W2}) = 2.7048 \times 10^{-6}$, $f(\mathrm{W3}) = 1.8326 \times 10^{-6}$, and $f(\mathrm{W4}) = 5.2269 \times 10^{-5}$.

\subsection{Error estimations}
\label{app:photo_err}
An estimate of the uncertainty for photometric fluxes can be computed as follows: 
\begin{equation}
    \sigma_\text{reg} = \sqrt{\sum_{i=1}^{N_\mathcal{A}}{(\sigma_i^2 + \frac{|f^\text{BGsub}_i|}{g_i})}},
    \label{eq:errorphoto}
\end{equation} where the region considered, $\mathcal{A}$, defines a set of $N_\mathcal{A}$ pixels $i$, corresponding to the aperture, $\sigma_i$ is the standard deviation of noise estimated from the background, in ADU (analog to digital units), $f^\text{BGsub}_i$ is the background-subtracted pixel value, and $g_i$ the effective gain of the detector in electron/ADU. The left term corresponds to detector noise, whereas the right term is associated with photon noise, considering the photon flux is following Poissonian statistics. When available -- for example in SDSS files header -- we added the dark variance $\sigma^2_{\text{dark},i}$, in ADU.

Concerning GALEX, we use eq. \ref{eq:errorphoto}. In GALEX files, the gain is taken to be $g=1$ as they mention "counts" to be detected photons. The intensity map (\texttt{-int.fits}) is in counts pix$^{-1}$ s$^{-1}$, so in order to handle solely counts, we computed uncertainties using background-subtracted, corrected for Galactic extinction count map (\texttt{-cnt.fits}). Only then we divided the uncertainties by the relative response file (\texttt{-rrhr.fits}), which is the result of sensitivity multiplied by the exposure time, to retrieve errors in counts pix$^{-1}$ s$^{-1}$.

Regarding uncertainty estimation for WISE files, we based our estimations on their documentation\footnote{\url{https://irsa.ipac.caltech.edu/data/WISE/docs/release/AllWISE/expsup/sec4_3a.html}} and our final estimate is
\begin{equation}
\label{eq:err_WISE}\sigma_\text{reg}^{WISE}=\sqrt{F_{corr}\left(\sum_i^{N_\mathcal{A}}\sigma_{unc}^2 + \frac{N_\mathcal{A}^2}{N_\text{BG}}\sigma^2_i \right)},
\end{equation}
with $F_{corr}$ is the correlated noise correction factor for flux variance, $\sigma_{unc}$ flux uncertainty for pixel i from uncertainty map, and $N_\text{BG}$ is the number of pixels in the area where background noise $\sigma_i$ is measured.

\subsection{Integrated fluxes}
See ~\ref{tab:photo_flux}. The corresponding SED fits to the listed fluxes integrated in the two regions considered (global: $4~R_e$, and local: PBCO) are shown in App~\ref{app:fits_globaux}.
\begin{table}[!htbp]
\caption{Flux density values in all bands convolved for the two global regions drawn in Fig. \ref{fig:CO+SDSS}.} 
\label{tab:photo_flux}
\centering
\begin{tabular}{lcc}
\hline\hline
Band & $F_{4Re}$ & $F_\t{PBCO}$ \\
     & {\small mJy} & {\small mJy}         \\
\hline
GALEX $FUV$ & 0.501 & 0.255 \\
GALEX $NUV$ & 0.737 & 0.381 \\
SDSS $u$    & 1.47 & 1.02 \\
SDSS $g$    & 5.50 & 3.84  \\
SDSS $r$    & 10.6 & 7.40  \\
SDSS $i$    & 15.6 & 10.9  \\
SDSS $z$    & 19.3 & 14.0  \\
WISE 1      & 21.3 & 15.3  \\
WISE 2      & 13.0 & 9.98  \\
WISE 3      & 58.8 & 43.7  \\
WISE 4      & 84.6 & 56.7  \\
\hline
\end{tabular}
\tablefoot{The results for the fits where these flux were used as priors can be found in Table \ref{tab:glob_prop}. The typical uncertainty is 10\%.}
\end{table}

\subsection{Galactic extinction}
See Table~\ref{tab:Gal_extinction}.
\begin{table}[h]
\caption{Galactic extinction values used for correcting the UV and optical data.}
\label{tab:Gal_extinction}
\centering
\begin{tabular}{lcc}
\hline\hline
Band & $\lambda$ & $A_\lambda$\\
& {\small Å} & {\small mag}         \\
\hline
GALEX $FUV$ & 1548 & 0.094\\
GALEX $NUV$ & 2303 & 0.101\\
SDSS $u$    & 3551 & 0.053\\
SDSS $g$    & 4686 & 0.042\\
SDSS $r$    & 6166 & 0.029\\
SDSS $i$    & 7480 & 0.020\\
SDSS $z$    & 8932 & 0.015\\
\hline
\end{tabular}
\end{table}

\onecolumn

\section{Integrated SED best fits}
\label{app:fits_globaux}

\begin{figure}[tph]
      \centering
      \begin{minipage}[b]{0.5\linewidth}
          \centering
          \includegraphics[height=7.5cm]{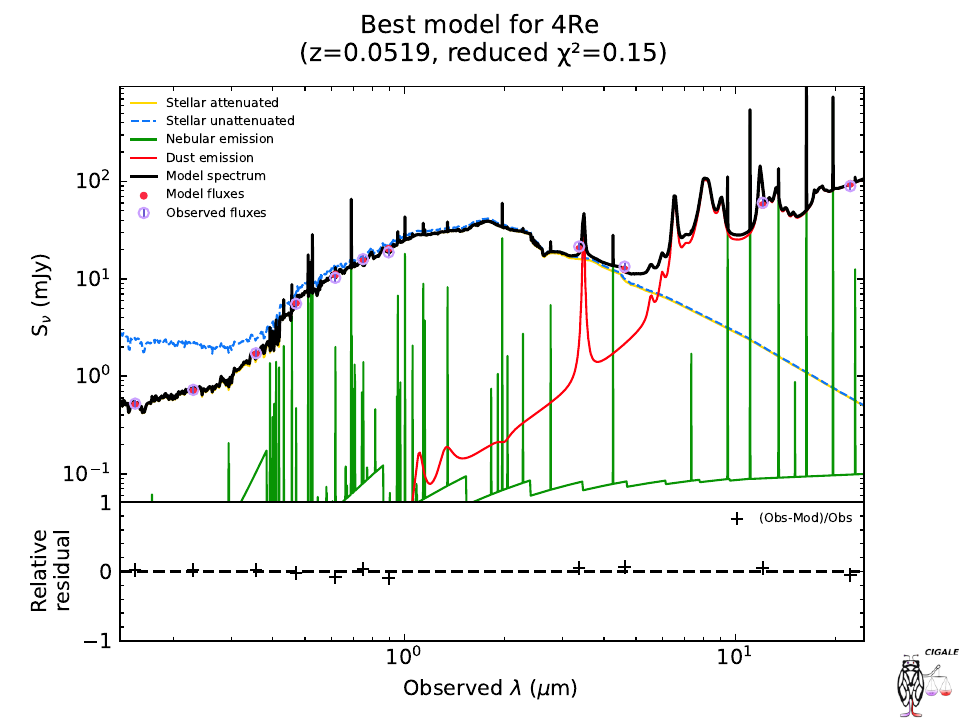}
      \end{minipage}%
      \begin{minipage}[b]{0.5\linewidth}
          \centering
          \includegraphics[height=7.5cm]{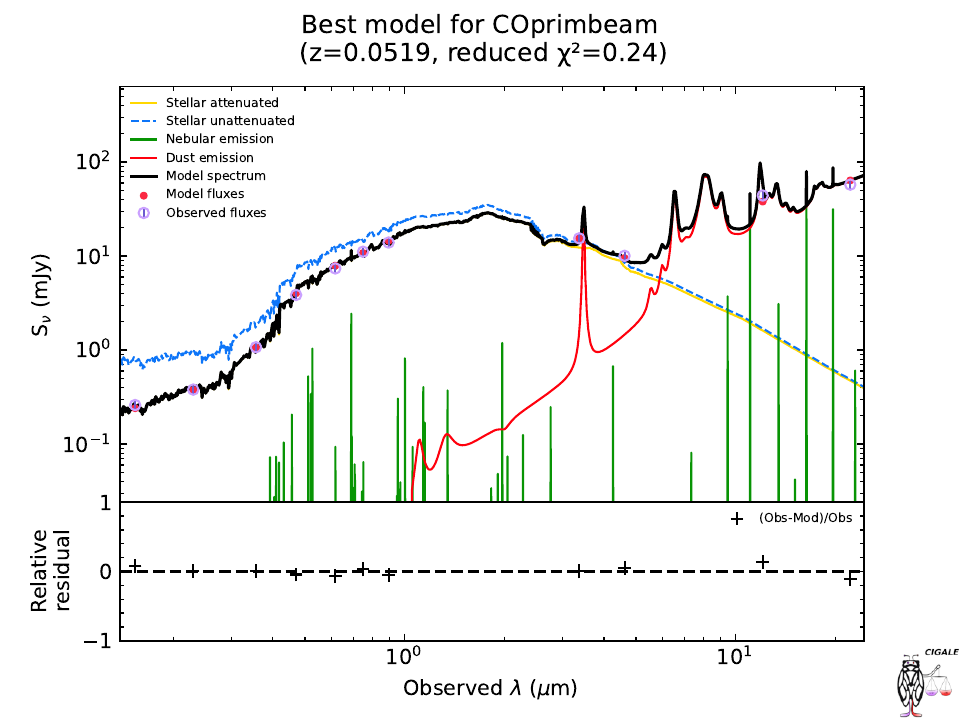}
      \end{minipage}
      \caption{\textit{Left:} Best SED fit for the total flux of UGC~8179, output by CIGALE. \textit{Right:} Best SED fit for the flux of UGC~8179, integrated in the region of CO detection and output by CIGALE.}
      \label{fig:fit_global}
\end{figure}

\section{Resolved SFMS double fit and rMGMS counterparts}

\begin{figure}[h!]
   \centering
   \includegraphics[width=\hsize]{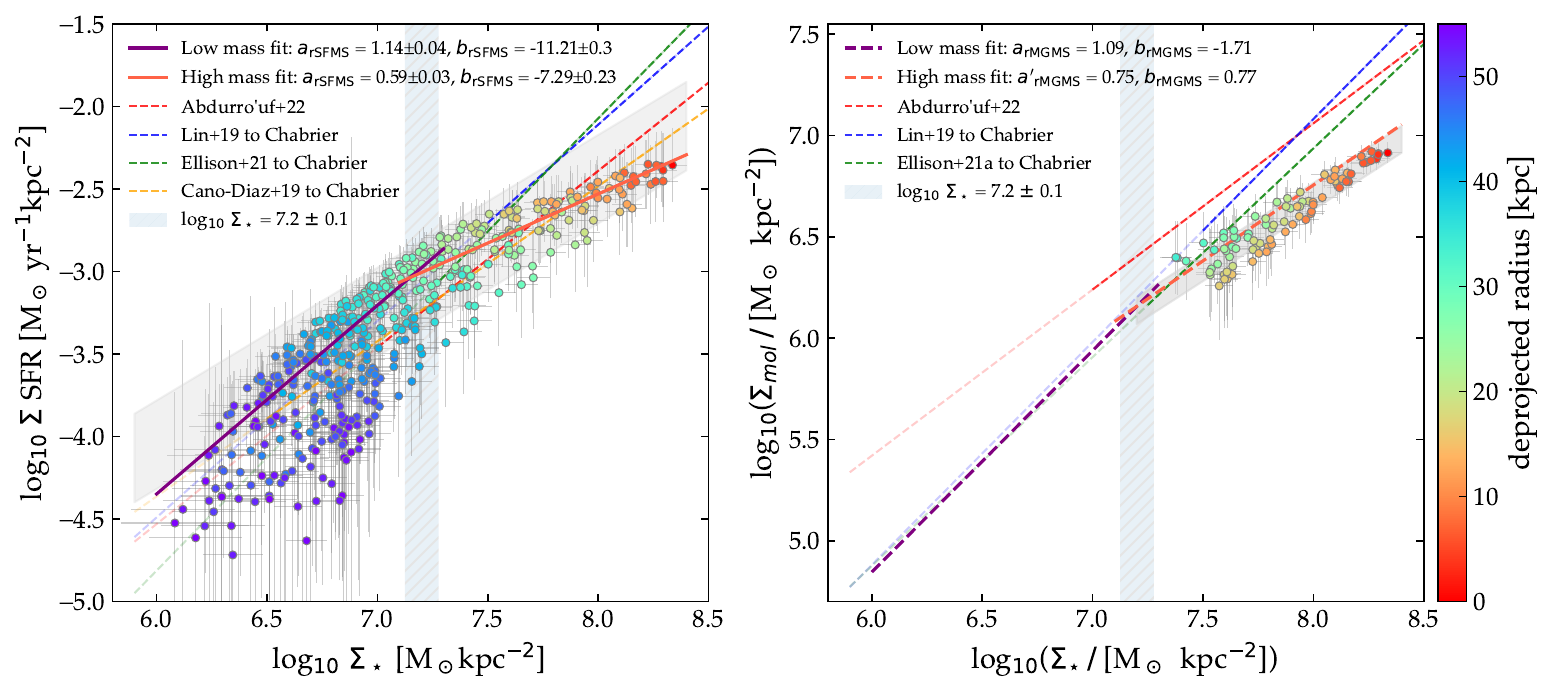}
    \caption{\textit{Left:} Resolved SFMS with two linear fits for high- and low-mass regimes respectively drawn as orange and purple solid lines. The hatched region corresponds to the range of values contributing both to the high- and low-mass ODR fits. The shaded area represents the scatter of the sole linear fit from the left panel of Fig.~\ref{fig:rSFMS_rKS_rMGMS}. \textit{Right:} Resolved MGMS with the two linear predictions from the opposite SFMS fits represented by the dotted orange and purple lines for the high- and low-mass regimes. The dotted purple line is a prediction of the rMGMS at low surface densities, below the molecular-to-atomic phase transition. The shaded area represents the scatter of the sole linear fit from the right panel of Fig.~\ref{fig:rSFMS_rKS_rMGMS}. The rest of the legend in both panels is identical to the left and right-hand panels of Fig.~\ref{fig:rSFMS_rKS_rMGMS}.}
    \label{fig:rSFMS_double_fit}
\end{figure}

\section{BPT diagrams}
\begin{figure}[h]
      \centering
      \begin{minipage}[b]{0.3\linewidth}
          \centering
          \includegraphics[height=5.5cm]{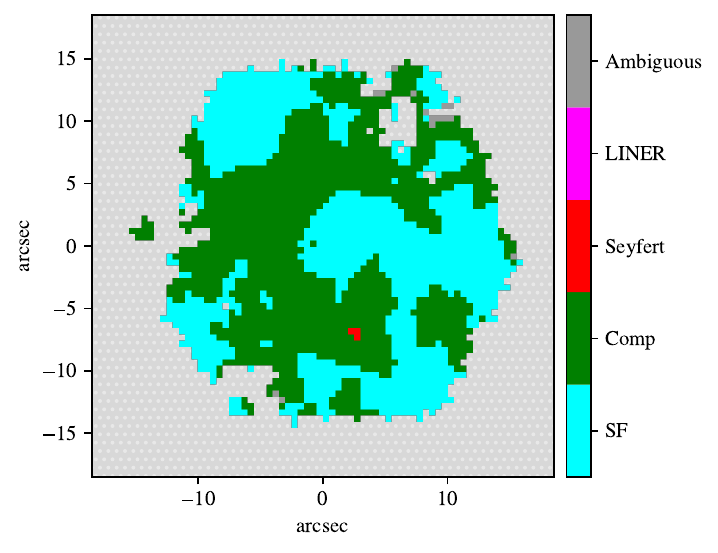}
      \end{minipage}
      \begin{minipage}[b]{0.6\linewidth}
          \centering
          \includegraphics[height=5.5cm]{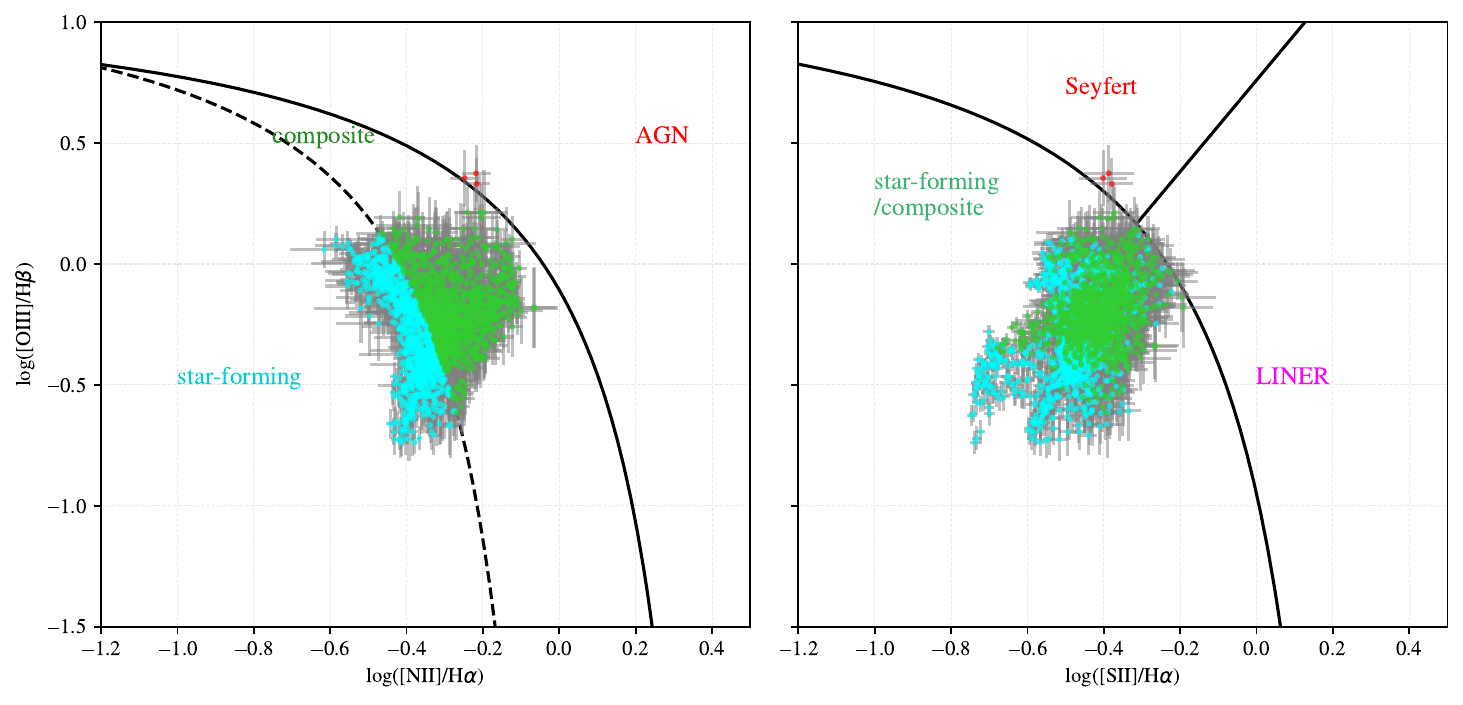}
      \end{minipage}
      \caption{\textit{Left:} Mapping of the BPT classification within the MaNGA FoV. \textit{Middle:} Corresponding NII-BPT diagram. \textit{Right:} Corresponding SII-BPT diagram.}
      \label{fig:BPT_MaNGA}
\end{figure}

Under the assumption of photoionization as prevalent ionization mechanism, we employed the emission line diagnostics known as BPT diagrams to classify star-forming, composite, LINER and Seyfert regions \citep[see][]{baldwin1981,kewley2001,kauffmann2003,kewley2006} 
To maximize spatial coverage of the classification, we used relaxed S/N thresholds (S/N(H$\alpha$)=3, S/N(H$\beta$)=2, S/N([NII])=3, S/N([SII])=2, S/N([OIII])=1) and omit the [OI]-based BPT diagram due to its limited added value and weak [OI] signal. We show the NII- and SII-BPT diagrams in Fig. \ref{fig:BPT_MaNGA}, along with a mapping of the corresponding classified pixels.

\section{Resolved KS with varying $\aCO$}

\begin{figure}[h!]
    \centering\includegraphics[width=\hsize]{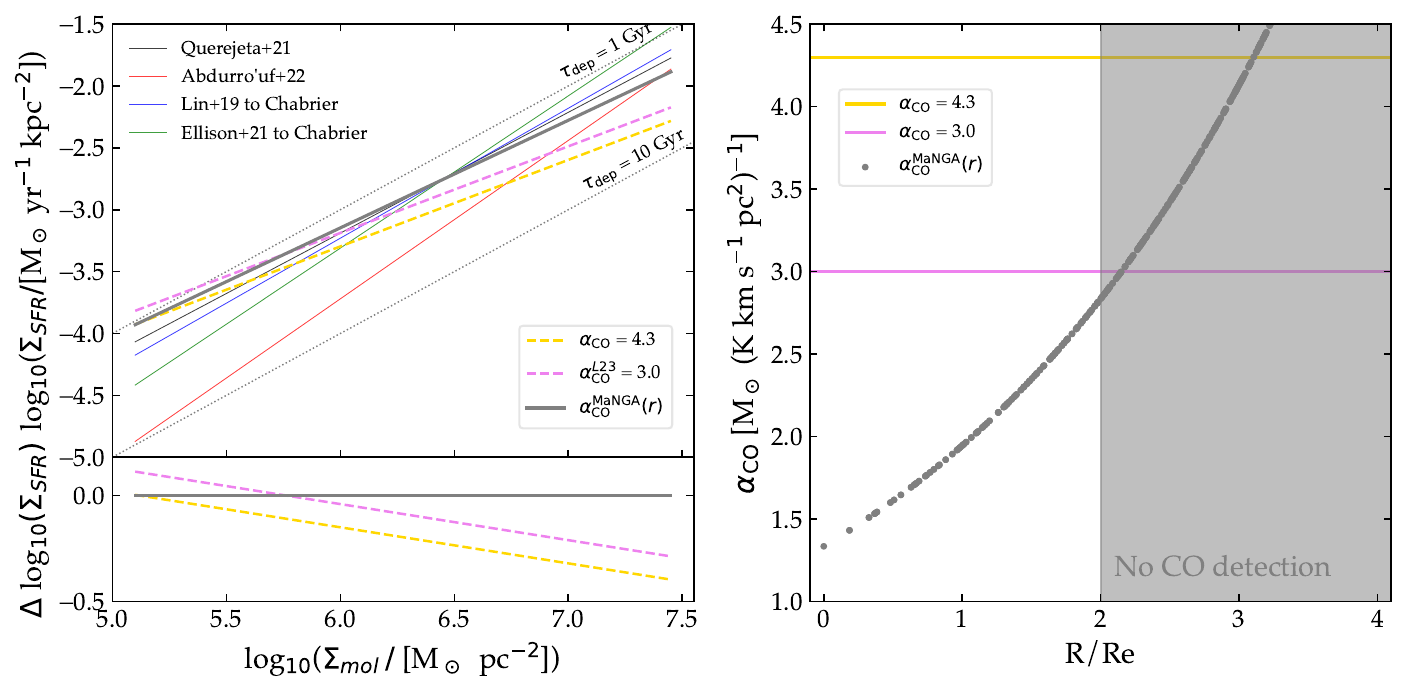}
    \caption{\textit{Top left}: Resolved Kennicutt-Schmidt relation with varying $\aCO$. The thick solid gray line corresponds to the ODR fit in the middle panel of Fig.~\ref{fig:rSFMS_rKS_rMGMS}, using a metallicity-dependent $\aCO^\t{MaNGA}$. The dashed pink (yellow) line represents the fit to the data computed using a constant $\aCO=3.0$ (4.3). The corresponding data points to each linear fit are only the unhatched pixels above the 4$\sigma$ detection threshold. \textit{Bottom left}: Residuals with respect to the results shown in Sec.~\ref{sec:results}, with $\aCO^\t{MaNGA}$. \textit{Right}: Radial profiles of $\aCO$ for the three prescriptions.}
    \label{fig:app:variation_KS_aCO}
\end{figure}
\clearpage
\section{Comparison studies}
\begin{table*}[h!]
    \begin{center}
    \caption{Sample properties from comparison studies.}
    \label{tab:compa_literature}
    \centering
    \begin{tabular}{lcccccc}
    \hline\hline
      Ref. & $z$ & \# of sources & Mass & Morpho. & Other citeria & $\aCO$\\
       & & & {\small\Msun} & & & {\tiny\Msun~(K~km~s$^{-1}$~pc$^{2}$)$^{-1}$}\\
    \hline 
    A22 & $0.001 \leq z \leq 0.005 $ & 10 & $10^{9.7} < \t{\M*} < 10^{11.2}$ & Spirals & e < 0.6 and b/a > 0.4 & 4.35\\
    E21\tablefootmark{a} & $0.02 < z < 0.05$ & 28 & $10^{10} < \t{\M*} < 10^{11.5}$ & SF spirals & & 4.3\\
    L19\tablefootmark{a} & $z \sim 0.03$ & 14 & $10^{10} < \t{\M*} < 10^{11.5}$ & SF spirals & $10^{-10.5} < \t{sSFR}< 10^{-9.5}$  & 4.3\\
    CD19\tablefootmark{b} & $0.01 < z < 0.15 $ & 1754 (459 LTGs) & $10^{9} < \t{\M*} < 10^{12}$ & all & i < 60° & -- \\
    Q21\tablefootmark{c} &$ z \lesssim 0.004 $ & 74 & $10^{9.25} < \t{\M*} < 10^{11.25}$ & SF spirals\tablefootmark{d} & & $4.35\times Z'^{-1.6}$\\
    \hline
    \end{tabular}
\tablefoot{
    \tablefoottext{a}{The parent sample is ALMaQUEST.}
    \tablefoottext{b}{The parent sample is MaNGA MPL-5.}
    \tablefoottext{c}{The parent sample is PHANGS.}
    \tablefoottext{d}{Q21 specify the presence of both late- and early-type spirals, along with a few lenticular (S0) and irregular galaxies.}}
    \end{center}
\end{table*}

\twocolumn
\end{document}